\documentclass[]{aa} 
\pdfoutput=1
\usepackage{graphicx}
\usepackage{txfonts}
\usepackage{natbib}
\usepackage{enumerate}
\usepackage[squaren,Gray]{SIunits}
\usepackage{multirow}
\usepackage{fixltx2e}
\usepackage{lscape}
\begin{document}

   \title{Hot Jupiters with relatives: discovery of additional planets in orbit around WASP-41 and WASP-47
   \thanks{Using data collected at ESO's La Silla Observatory, Chile: HARPS on the ESO 3.6m (Prog ID 087.C-0649 \& 089.C-0151), the Swiss {\it Euler} Telescope, TRAPPIST, the 1.54-m Danish telescope (Prog CN2013A-159), and at the LCOGT's Faulkes Telescope South. The data is publicly available at the \textit{CDS} Strasbourg and on demand to the main author.}
   }

\titlerunning{Hot Jupiters with relatives}

   \author{M. Neveu-VanMalle\inst{1,2}
                \and
                D. Queloz\inst{2,1}
                \and
                D. R. Anderson\inst{3}
                \and
                D. J. A. Brown\inst{4}
                \and
                A. Collier Cameron\inst{5}
                \and
                L. Delrez\inst{6}
                \and
                R. F. D\'iaz\inst{1}
                \and
                M. Gillon\inst{6}
                \and
                C. Hellier\inst{3}
                \and
                E. Jehin\inst{6}
                \and
                T. Lister\inst{7}
                \and
                F. Pepe\inst{1}
                \and
                P. Rojo\inst{8}
                \and
                D. S\'egransan\inst{1}
                \and
                A. H. M. J. Triaud\inst{9,10,1}
                \and
                O. D. Turner\inst{3}
                \and
                S. Udry\inst{1}
          }

   \institute{Observatoire Astronomique de l'Universit\'e de Gen\`eve,
                Chemin des Maillettes 51, 1290 Sauverny, Switzerland
          \and
             Cavendish Laboratory, J J Thomson Avenue, Cambridge, CB3 0HE, UK
             \and
             Astrophysics Group, Keele University, Staffordshire, ST5 5BG, UK
             \and
             Department of Physics, University of Warwick, Gibbet Hill Road, Coventry CV4 7AL, UK
             \and
             SUPA, School of Physics and Astronomy, University of St. Andrews, North Haugh,  Fife, KY16 9SS, UK
             \and
             Institut d'Astrophysique et de G\'eophysique, Universit\'e de
            Li\`ege, All\'ee du 6 Ao\^ut, 17, Bat. B5C, Li\`ege 1, Belgium
             \and
             Las Cumbres Observatory Global Telescope Network, 6740 Cortona Dr. Suite 102, Goleta, CA 93117, USA
              \and
              Departamento de Astronom\'ia, Universidad de Chile, Camino El Observatorio 1515, Las Condes, Santiago, Chile
              \and
             Centre for Planetary Sciences, University of Toronto at Scarborough, 1265 Military Trail, Toronto, ON, M1C 1A4, Canada
             \and
             Department of Astronomy \& Astrophysics, University of Toronto, Toronto, ON, M5S 3H4, Canada
             \\
             \email{marion.neveu@unige.ch}
             }

   \date{Received July 15, 2015; accepted September 30, 2015}

 
  \abstract
  {We report the discovery of two additional planetary companions to WASP-41 and WASP-47. WASP-41 c is a planet of minimum mass 3.18 $\pm$ 0.20 M$_{\rm Jup}$ and eccentricity 0.29 $\pm$ 0.02, and it orbits in 421 $\pm$ 2 days. WASP-47 c is a planet of minimum mass 1.24 $\pm$ 0.22 M$_{\rm Jup}$ and eccentricity 0.13 $\pm$ 0.10, and it orbits in 572 $\pm$ 7 days. Unlike most of the planetary systems that include a hot Jupiter, these two systems with a hot Jupiter have a long-period planet located at only $\sim$1\,au from their host star. WASP-41 is a rather young star known to be chromospherically active. To differentiate its magnetic cycle from the radial velocity effect induced by the second planet, we used the emission in the H$\alpha$ line and find this indicator well suited to detecting the stellar activity pattern and the magnetic cycle. The analysis of the Rossiter--McLaughlin effect induced by WASP-41 b suggests that the planet could be misaligned, though an aligned orbit cannot be excluded. WASP-47 has recently been found to host two additional transiting super Earths. With such an unprecedented architecture, the WASP-47 system will be very important for understanding planetary migration.}

              
   \maketitle
%

\section{Introduction}

In contrast to the large number of multiple planetary systems, stars with hot-Jupiter planets have long been thought to lack additional planets.
Recent studies suggest that this statement may not be true \citep[e.g.][]{Knutson:2014lr}. However the existence of those additional planets was revealed by nothing more than radial velocity trends. Only seven outer planetary companions of close-in (a < 0.1\,AU) giant planets have a full orbital period that has been observed. 

The multiple planetary system around \object{$\upsilon$ Andromedae} \citep{Butler:1997fk} was the first to be found with a hot Jupiter. This planet is surrounded by three more massive giant planets orbiting between 0.8 and 5 au \citep{Curiel:2011qy}. Ten years later, the discovery of the planetary system around \object{HIP 14810} \citep{Wright:2007uq, Wright:2009fj} revealed that a hot Jupiter can be found in a system with smaller planets on wider orbits. From 15\,years of radial velocity surveys, \citet{Feng2015} recently constrained the orbits of the distant (P > 9\,years) giant planets in orbit around the hot-Jupiter hosts \object{HD 187123} and \object{HD 217107}. Among all systems with a close-in giant planet discovered by transit surveys, only \object{HAT-P-13} \citep{Bakos:2009kx}, \object{HAT-P-46} \citep{Hartman:2014yq}, and \object{Kepler-424} \citep{Endl:2014vn} have an additional planetary companion detected by Doppler surveys with a fully measured orbit.

The Wide Angle Search for Planets \citep[WASP,][]{Pollacco:2006qy} has discovered more than 100 hot Jupiters. While some of them clearly have a very distant companion \citep[e.g. WASP-8,][]{Queloz:2010lr}, none of them were previously known to have a fully resolved orbit.

The radial velocity follow-up observations of WASP initiated in 2008 with the fibre-fed echelle spectrograph CORALIE (mounted on the 1.2-m Euler Swiss telescope at La Silla) has identified new multiple systems including a hot Jupiter. We report here the detection of additional planets around WASP-41 \citep{Maxted:2011fk} and WASP-47 \citep{Hellier:2012uq}. These new planets are both located only $\sim$1\,au of their parent star.
When searching for long-period companions using the radial velocities technique, the effects of stellar activity should be carefully considered to avoid misdetections. A stellar magnetic cycle may mimic the signal of a planetary companion \citep[e.g.][]{Lovis:2011lr}. WASP-41 is known to be active \citep{Maxted:2011fk}, which makes it important to recognise the temporal structure of its magnetic cycle. Since the photospheric Ca (H \& K) emission lines cannot be extracted from single CORALIE spectra (insufficient signal-to-noise ratio), we alternatively use the emission in the H$\alpha$ band to characterise activity. We find that this indicator is very well suited to WASP-41, though it may not be true in general \citep{Cincunegui:2007lr}.

Section \ref{w47c} reports the observations, analysis, and results for WASP-47. Section \ref{w41c} similarly addresses the case of WASP-41 with a description of the various additional observations, a study of the stellar activity, and an analysis that includes the Rossiter--McLaughlin effect followed by the results. In section \ref{disc}, after discussing the importance of considering magnetic cycles when searching for long-period planets, we compare our two multiple planetary systems, including a hot Jupiter, to the five previously known ones. We discuss the biases induced by the observing strategies of Doppler and transit surveys in finding multiple planetary systems including a hot Jupiter. Finally we check the transit probabilities of \object{WASP-47c} and \object{WASP-41c}.

During the process of revision of this paper, \citep{Becker:2015lr} reported the presence of two additional super-Earths transiting WASP-47 every $\sim$0.8 and 9 days. This remarkable discovery brings strong constraints on the migration of this system and will be discussed in the last section.


\section{WASP-47 c} \label{w47c}

\subsection{Observations}
\begin{figure}
   \centering
   \resizebox{\hsize}{!}{\includegraphics{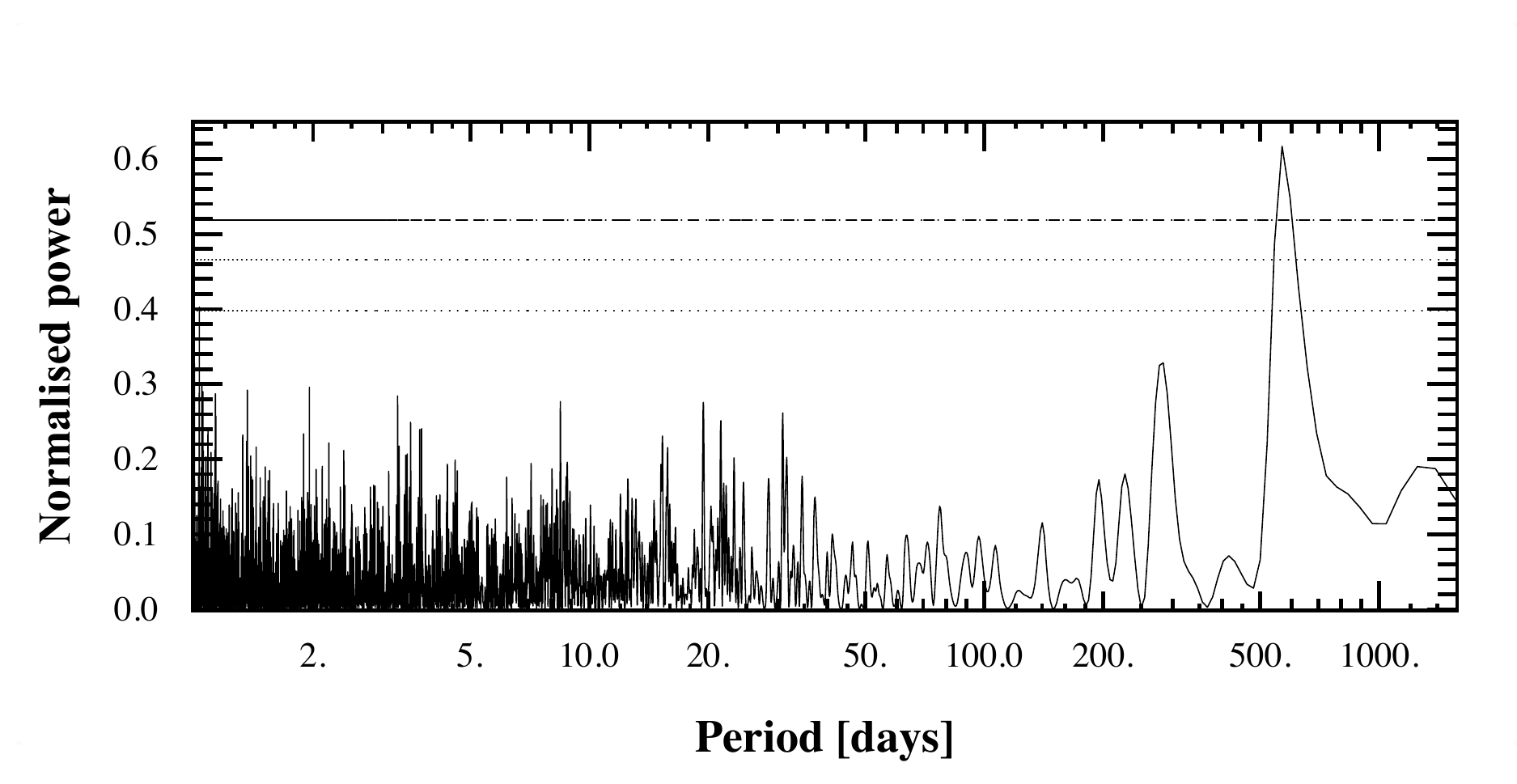}}
      \caption{WASP-47 Lomb-Scargle periodogram of the residuals after subtracting Planet b. The dotted lines correspond to a false-alarm probability of 0.1\%, 1\%, and 10\%.}
         \label{per_w47}
   \end{figure}

WASP-47 is a G9V star hosting a giant planet with a mass of 1.14\,M$_{\rm Jup}$ and a period of 4.16 days \citep{Hellier:2012uq}. The discovery paper includes 19 radial velocity measurements taken with CORALIE during two observing seasons. Ongoing monitoring revealed a possible third body, and after 46 observations over 4.44 years, the orbit of a second planet was clear. 
After subtracting the inner planet (hot Jupiter) from the radial velocity data, the analysis of the residuals using a Lomb-Scargle periodogram shows a clear peak around 550 days (see Fig. \ref{per_w47}). The analysis of the residuals after subtracting the two-planet model shows no additional signal (see Fig. \ref{w47_ap}).

In November 2014, an upgrade of CORALIE was performed. The changes introduced an offset in the radial velocities. Since the orbit of the outer planet was not fully covered in phase, we collected six additional measurements from June to August 2015. One measurement was done right after the upgrade just before the star became unobservable but it suffered from daylight contamination and had to be discarded. For the analysis, the 2015 data are considered as an independent radial velocity data set.

\subsection{Analysis}
We combined the photometry data from EulerCAM (Gunn-$r'$ filter) published in \citet{Hellier:2012uq} with all the radial velocities from CORALIE and used an adaptive Markov chain Monte-Carlo (MCMC) fitting scheme to derive the system parameters of the two-planet model. We did not include the WASP photometry in our analysis because it has poor precision compared to EulerCAM and might suffer from spatial dilution from background stars. We used the most recent version of the code described in \citet{Gillon:2012lr} and references therein. The jump parameters used in our analysis to characterise the transiting planet \object{WASP-47b} were
\begin{itemize} 
\item the transit depth defined as the planet/star area ratio (d$F=(R_{\rm p}/R_{\star})^{2}$), where $R_{\rm p}$ and $R_{\star}$ are the planetary and stellar radius, respectively;
\item the transit impact parameter in the case of a circular orbit ($b'=a_{\rm b}\cos i_{\rm p,b}/R_{\star}$), where $a_{\rm b}$ and $i_{\rm p,b}$ are the semi-major axis and the inclination of the planetary orbit, respectively;
\item the transit duration $T_{\rm 14}$, from first to last contact;
\item the orbital period $P_{\rm b}$; 
\item the time of inferior conjunction $T_{\rm 0,b}$;
\item the two parameters $\sqrt{e_{\rm b}}\,\cos\omega_{\rm b}$ and $\sqrt{e_{\rm b}}\,\sin\omega_{\rm b}$, where $e_{\rm b}$ is the eccentricity and $\omega_{\rm b}$ the argument of periastron;
\item the parameter $K_{\rm 2,b}=K_{\rm b}\sqrt{1-e_{\rm b}^{2}}\,P_{\rm b}^{1/3}$, where $K_{\rm b}$ is the radial velocity semi-amplitude.
\end{itemize}
For the period $P_{\rm b}$ and time of inferior conjunction $T_{\rm 0,b}$, we imposed Gaussian priors defined by the values derived in \citet{Hellier:2012uq}, since we did not include the WASP photometry in our analysis. Uniform priors were assumed for the probability distribution functions of all the other jump parameters. The choices of jump parameters were optimised to reduce the dependencies between the parameters and thus to minimise the correlations. We checked that using $\sqrt{e}\,\cos\omega$ and $\sqrt{e}\,\sin\omega$ as jump parameters translates into a uniform prior on e and does not influence the results.

We modelled\ the limb-darkening with a quadratic law, where instead of the coefficients $u_{1}$ and $u_{2}$, we used the combinations $c_{1}=2\times u_{1}+u_{2}$ and $c_{2}=u_{1}-2\times u_{2}$ as jump parameters in order to minimise the correlation of the obtained uncertainties \citep{Holman:2006fk}. We assumed normal priors on $u_{1}$ and $u_{2}$ with the values deduced from the tables by \citet{Claret:2011qy}.
As described in \citet{Gillon:2012lr}, we applied a correction factor $CF=1.32$ to the error bars of the photometric data to account for the red noise.

WASP-47 c does not have transit observations, the jump parameters used in our analysis to define the second orbit were
\begin{itemize} 
\item the orbital period $P_{\rm c}$; 
\item the time of inferior conjunction $T_{\rm 0,c}$;
\item the two parameters $\sqrt{e_{\rm c}}\cos\omega_{\rm c}$ and $\sqrt{e_{\rm c}}\sin\omega_{\rm c}$;
\item the parameter $K_{\rm 2,c}=K_{\rm c}\sqrt{1-e_{\rm c}^{2}}\,P_{\rm c}^{1/3}$.
\end{itemize}
We assumed uniform priors on all those jump parameters.

The stellar density is derived at each step of the MCMC from Kepler's third law and the values of the scale parameter $a_{\rm b}/R_{\star}$ and orbital period of WASP-47b \citep{Winn:2010lr}.
The stellar mass $M_\star$ is obtained thanks to a calibration \citep{Enoch:2010uq,Gillon:2011fj} from well-constrained binary systems \citep{Southworth:2011kx}. This empiric law is a function of $T_{\rm eff}$, $\rho_{\star}$ and [Fe/H]. We used Gaussian prior distributions for $T_{\rm eff}$ and [Fe/H], and the initial value for $\rho_{\star}$, based on the values derived by \citet{Mortier:2013yq}.

\subsection{Results}
\begin{figure}
   \centering
   \includegraphics[width=10cm]{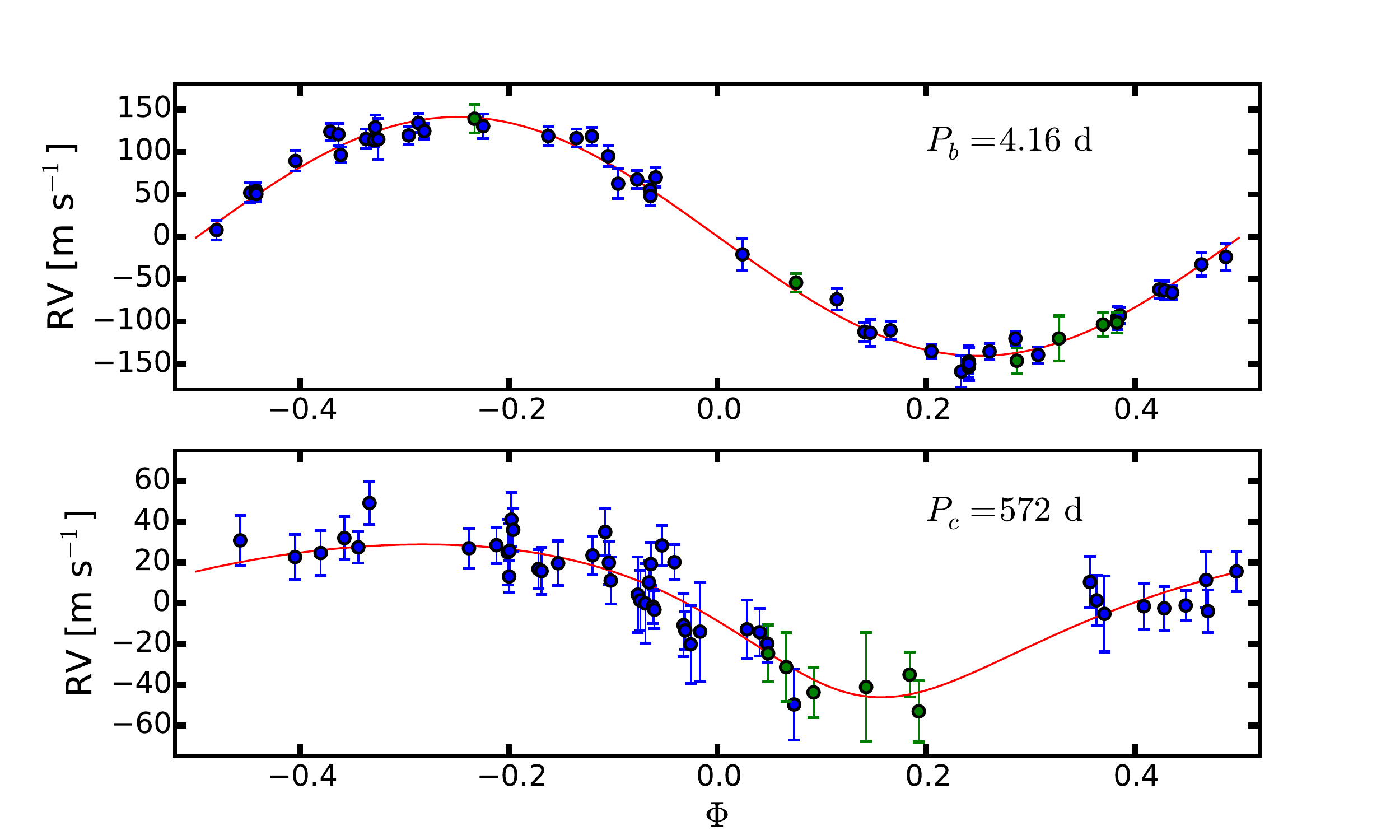}
      \caption{WASP-47 CORALIE radial velocity data (blue/green dots: before/after the upgrade) and best fit model (red line). Top:  phase-folded on the period of the inner planet (outer planet subtracted). Bottom: phase-folded on the period of the outer planet (inner planet subtracted).}
         \label{plot_w47}
   \end{figure}

The results presented in Table \ref{res_w47} have been obtained running five chains of 100\,000 accepted steps. The derived values are the median and 1 $\sigma$ limits of the marginalised posterior distributions, considering a 20\% burn-in phase. The acceptance fraction is $\sim$25\% ($\sim$40\% for the burn-in phase). We used the statistical test of \citet{Gelman92} to check the convergence of the chains and get a potential scale reduction factor (PSRF) smaller than 1.01 for all the jump parameters. The autocorrelations are shorter than 2\,000 steps except for the semi-amplitude and the eccentricity of the outer planet. The lack of data around the radial velocity minimum introduces a small degeneracy between these parameters. Their autocorrelations are $\sim$5\,000 steps long.
As we did not include additional transit light curves, we did not improve the determination of the transit parameters compared to \citet{Hellier:2012uq}. They are not given here
for this reason. The best fit solution for the radial velocities is plotted in Fig. \ref{plot_w47}. The values provided in Table \ref{res_w47} are the median values of the posterior distributions and not the best fit values used in Fig. \ref{plot_w47}. To verify that the median values represent a self-consistent set of parameters, we compare the BIC (Bayesian Information Criterion) obtained using the median values and the best fit values. We obtain compatible results.

Including the data collected in 2015 allows to reduce the uncertainty on the eccentricity and semi-amplitude of the outer planet. All the other parameters remain unchanged compared to the results obtained with the first data set alone. The posterior distribution of the offset between the two data sets ($\Delta$RV = 18 $\pm$ 11\,m s$^{-1}$) agrees with what has been observed on bright radial velocity standards.

The semi-amplitude of WASP-47 c is barely twice as large as the radial velocity uncertainties, and the phase coverage is incomplete. To validate the detection, we compared the results obtained when running the same MCMC with a model including only one planet. With the one-planet model, the reduced $\chi ^{2}$ on the initial set of radial velocities is 2.7 and the residual r.m.s. 19.2\,m s$^{-1}$. When fitting two planets, the corresponding reduced $\chi ^{2}$ is 1.2 and the residual r.m.s. 10.6\,m s$^{-1}$. The two-planet model is clearly preferable.

\begin{table}
\centering
\caption{Median and 1-$\sigma$ limits of the posterior marginalised PDFs
obtained for the WASP-47 system derived from our global MCMC
analysis}
\small
\begin{tabular}{lc}
\hline\hline
\multicolumn{2}{l}{\bf MCMC Jump parameters}\\
\hline
\multicolumn{2}{l}{\bf Star}\\
$T_{\rm eff}$ [K] & 5576 $\pm$ 68\\
{[Fe/H]} & 0.36 $\pm$ 0.05\\
$c_{1}$ & 1.17 $\pm$ 0.03\\
$c_{2}$ & $-0.02 \pm$ 0.02\\
\multicolumn{2}{l}{\bf Planet b}\\
$P_{\rm b}$ [d] & 4.1591409 $\pm$ 0.0000072\\
$T_{\rm 0,b}$ -- 2\,450\,000 [BJD$_{\rm TDB}$] & 5764.3463 $\pm$ 0.0002\\$\sqrt{e_{\rm b}}\,\cos\omega_{\rm b}$  & 0.04 $\pm$ 0.06\\
$\sqrt{e_{\rm b}}\,\sin\omega_{\rm b}$ & 0.01 $\pm$ 0.09\\
$K_{\rm 2,b}=K_{\rm b}\sqrt{1-e_{\rm b}^{2}}\,P_{\rm b}^{1/3}$ [m s$^{-1}$ d$^{1/3}$] & 225.7 $\pm$ 3.6\\
\multicolumn{2}{l}{\bf Planet c}\\
$P_{\rm c}$ [d] & 572 $\pm$ 7\\
$T_{\rm 0,c}$ -- 2\,450\,000 [BJD$_{\rm TDB}$] & 5981 $\pm$ 16\\
$\sqrt{\rm e_{c}}\,\cos\omega_{\rm c}$  & $-0.21$ $\pm$ 0.24\\
$\sqrt{\rm e_{c}}\,\sin\omega_{\rm c}$ & 0.16 $\pm$ 0.20\\
$K_{\rm 2,c}=K_{\rm c}\sqrt{1-e_{\rm c}^{2}}\,P_{\rm c}^{1/3}$ [m s$^{-1}$ d$^{1/3}$] & 248 $\pm$ 43\\
\hline
\multicolumn{2}{l}{\bf Derived parameters}\\
\hline
\multicolumn{2}{l}{\bf Star}\\
$u_{1}$ & 0.462$\pm$0.014\\
$u_{2}$ & 0.242$\pm$0.010\\
Density $\rho_{\star}$ [$\rho_{\sun}$] & 0.68 $\pm$ 0.06\\
Surface gravity $\log g_{\star}$ [cgs] & 4.32 $\pm$ 0.05\\
Mass $M_{\star}$ [$M_{\sun}$] & 1.026 $\pm$ 0.076\\
Radius $R_{\star}$ [$R_{\sun}$] & 1.15 $\pm$ 0.04\\
Luminosity $L_{\star}$ [$L_{\sun}$] & 1.14 $\pm$ 0.11\\
\multicolumn{2}{l}{\bf Planet b}\\
Mass $M_{\rm p,b}$ [M$_{\rm Jup}$] & 1.13 $\pm$ 0.06\\
RV amplitude $K_{\rm b}$ [m s$^{-1}$] &140 $\pm$ 2\\
Orbital eccentricity $e_{\rm b}$ & < 0.026 (at 2\,$\sigma$)\\
Orbital semi-major axis $a_{\rm b}$ [au]& 0.051 $\pm$ 0.001\\
Density $\rho_{\rm p,b}$ [$\rho_{\rm J}$] & 0.71 $\pm$ 0.08\\
Surface gravity $\log g_{\rm p,b}$ [cgs] & 3.33 $\pm$ 0.03\\
Equilibrium temperature $T_{\rm eq,b}^{(*)}$ [K] & 1275 $\pm$ 23\\
\multicolumn{2}{l}{\bf Planet c}\\
Minimum mass $M_{\rm p,c}\sin i_{\rm p,c}$ [M$_{\rm Jup}$] & 1.24 $\pm$ 0.22\\
RV amplitude $K_{\rm c}$ [m s$^{-1}$] & 30 $ \pm$ 6\\
Orbital eccentricity $e_{\rm c}$ & 0.13 $\pm$ 0.10\\
Argument of periastron $\omega_{\rm c}$ [\degr] & 144 $\pm$ 53\\
Orbital semi-major axis $a_{\rm c}$ [au]& 1.36 $\pm$ 0.04\\
Equilibrium temperature $T_{\rm eq,b}^{(*)}$ [K] & 247 $\pm$ 5\\
$a_{\rm c}/R_{\star}$ & 255 $\pm$ 7\\
\hline
\end{tabular}
\label{res_w47}
\tablefoot{$^{(*)}$Assuming zero albedo and efficient redistribution of energy.}
\end{table}

\section{WASP-41 c} \label{w41c}
\subsection{Observations}
\begin{figure}
   \centering
   \resizebox{\hsize}{!}{\includegraphics{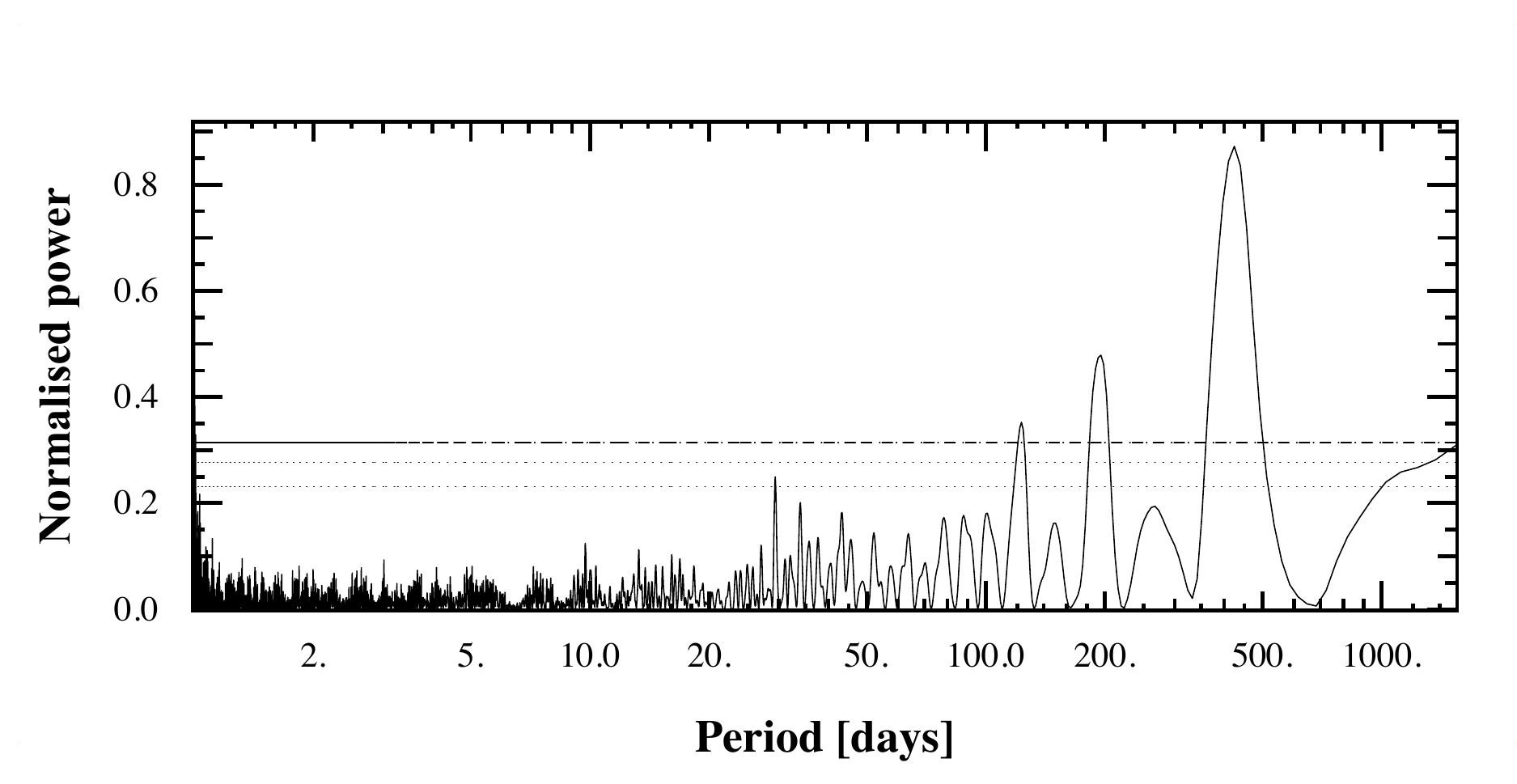}}
      \caption{WASP-41 Lomb-Scargle periodogram of the residuals after subtracting Planet b. The dotted lines correspond to false-alarm probabilities of 0.1\%, 1\%, and 10\%.}
         \label{per_w41}
   \end{figure}

WASP-41 is a G8V star hosting a giant planet with a mass of 0.9\,M$_{\rm Jup}$ and a period of 3.05 days \citep{Maxted:2011fk}. The discovery paper includes 22 radial velocity measurements taken within four months (excluding the first point). The short time span of the observations did not allow for detecting any additional signal. We did a follow-up on the star and collected a total of 100 radial velocities with CORALIE over 4.46 years. The observations during the second season revealed an offset of $\sim$100\,m\,s$^{-1}$ and triggered an intensive monitoring of the target. An additional planet on an eccentric orbit
rapidly became apparent. Because of the significant eccentricity of the second planet combined with a period close to one year, we had to wait until the fifth year of follow-up to plan observations at the periastron with a good sampling. After subtracting the inner planet (hot Jupiter) from the radial velocity data, the analysis of the residuals using a Lomb-Scargle periodogram shows a clear peak around 400 days (see Fig. \ref{per_w41}). The additional peaks present at $\sim$200 days and $\sim$130 days correspond to the two first harmonics $P/2$ and $P/3$. Such features are expected when applying frequency decomposition to eccentric signals, and WASP-41c has an eccentricity $e\sim0.3$. The analysis of the residuals, after subtracting the two planets from the radial velocities, shows no additional periodic signal (see Fig. \ref{w41_ap}).

We used the HARPS spectrograph mounted on the 3.6-m telescope in La Silla to observe the Rossiter--McLaughlin effect induced by WASP-41 b during the transit on 3 Apr 2011. In addition to the Rossiter--McLaughlin observations, 30 measurements out of transit of WASP-41 were gathered with HARPS between Apr 2011 and Aug 2012 to confirm the second signal detected with CORALIE.

More transit light curves of WASP-41 b have been obtained since the publication of the discovery paper: with the Faulkes Telescope South (FTS) at Siding Spring Observatory and with the TRAPPIST telescope \citep{Jehin:2011fj} and the Danish telescope both in La Silla. Details of these observations can be found in Table \ref{trans}.

   \begin{table}
   \centering
      \caption[]{Additional transit observations of WASP-41b and applied correction factors.}
         \begin{tabular}{l c c c}
            \hline\hline
            Instrument       &   Filter & date & $CF$\\
            \hline
            FTS  & PanSTARRS $z$ & 12 Apr 2011 & 1.8\\
            \hline
            \multirow{5}{*}{TRAPPIST} & \multirow{5}{*}{$I+z$} & 21 Mar 2011 & 1.0\\
            & & 02 Apr 2011 & 2.1\\
            & & 12 May 2011 & 2.0\\
            & & 09 Mar 2012 & 1.5\\
            & & 19 Apr 2013 & 1.4\\
            \hline
            \multirow{2}{*}{Danish} & \multirow{2}{*}{$R$} & 19 Apr 2013 & 2.2\\
            & & 25 Apr 2013 $^{(*)}$ & 3.0\\
            \hline
         \end{tabular}
    \label{trans}
    \tablefoot{$^{(*)}$ Partial transit}
   \end{table}

\subsection{Study of stellar activity}
WASP-41 is known to be an active star \citep{Maxted:2011fk}. In spite of the large amplitude of the second signal, we check that it is not due to a stellar magnetic cycle. For that purpose we searched for correlations between the radial velocities and spectral activity indicators. No variation is observed in the bisector span of the cross correlation function (Fig. \ref{w41_ap} bottom panel). The full width at half maximum (FWHM) shows some variability, which is suspiciously similar to the radial velocities' variability (periodicity close to 400 days). The $\log R'_{\rm HK}$ extracted from the HARPS spectra, based on the photospheric Ca II (H \& K) lines, indicates a mean value of $-4.483\pm0.036$ characteristic of an active star. But no variation in the $\log R'_{\rm HK}$ is visible during the two years of HARPS data (Fig. \ref{w41_rhk}).

The emission in the H$\alpha$ line is a rather underused activity indicator to trace photospheric stellar activity. In their study, \citet{Cincunegui:2007lr} concluded that the activity measured in the H$\alpha$ was not equivalent to the $R'_{\rm HK}$ indicator. They did not measure similar correlations between the two indicators in their whole sample. Some of their stars showed correlations, when others showed no or anti-correlations. Using solar observations, \citet{Meunier:2009fk} demonstrated that the Ca II (H \& K) and the H$\alpha$ emissions have different behaviours. During an active period when the filaments reach saturation, they observe a positive correlation between the indicators. On the other hand, during moderate active periods, the filaments counteract the effect of plages on the H$\alpha$ measurements, leading to weak or negative correlations. \citet{Gomes-da-Silva:2014qy} conclude that active stars with a mean $\log R'_{\rm HK} \geq -4.7$ systematically show a positive correlation between the Ca II (H \& K) and H$\alpha$ flux.

Since WASP-41 is a very active star with $\log R'_{\rm HK} \geq -4.7$, it is reasonable to use H$\alpha$ as an activity indicator. The motivation to use this band is that it is located in the red part of the stellar spectrum where a high signal-to-noise ratio can be obtained from the CORALIE spectra. As suggested in \citet{Cincunegui:2007lr}, we measured the H$\alpha$ emission at 6562.808\,$\angstrom$ and the continuum centred at 6605\,$\angstrom$ and averaged on a window 20\,$\angstrom$ wide. Unlike \citet{Cincunegui:2007lr} we extracted the H$\alpha$ from a 0.6\,$\angstrom$ wide band instead of 1.5\,$\angstrom$.

\begin{figure}
   \centering
      \includegraphics[width=10cm]{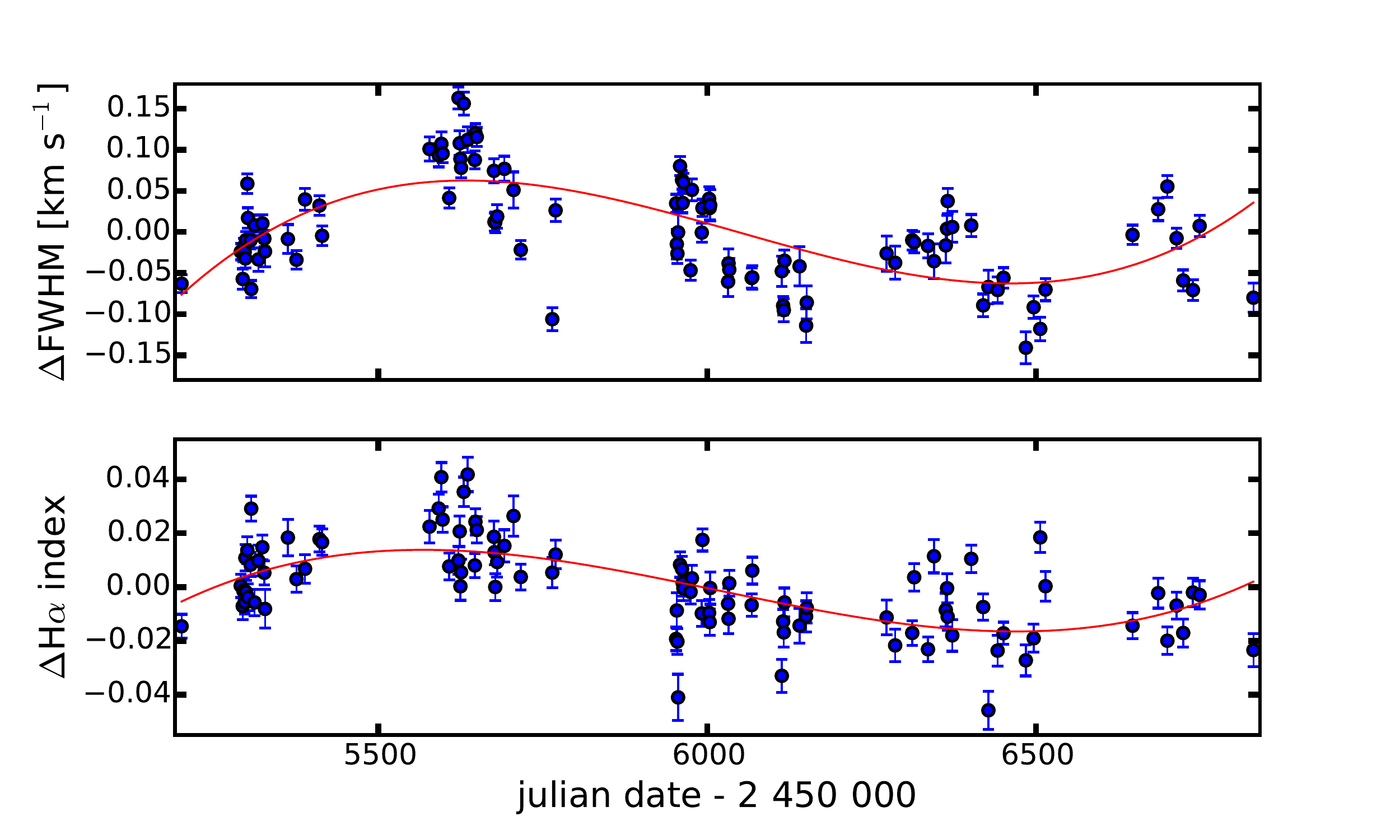}
      \caption{Magnetic cycle of WASP-41. Top: FWHM of the CCF from the CORALIE measurements. Bottom: H$\alpha$ index extracted from the CORALIE spectra. The red lines correspond to the fitted third-order trend.}
         \label{act_w41}
   \end{figure}

\begin{figure}
   \centering
   \resizebox{\hsize}{!}{\includegraphics{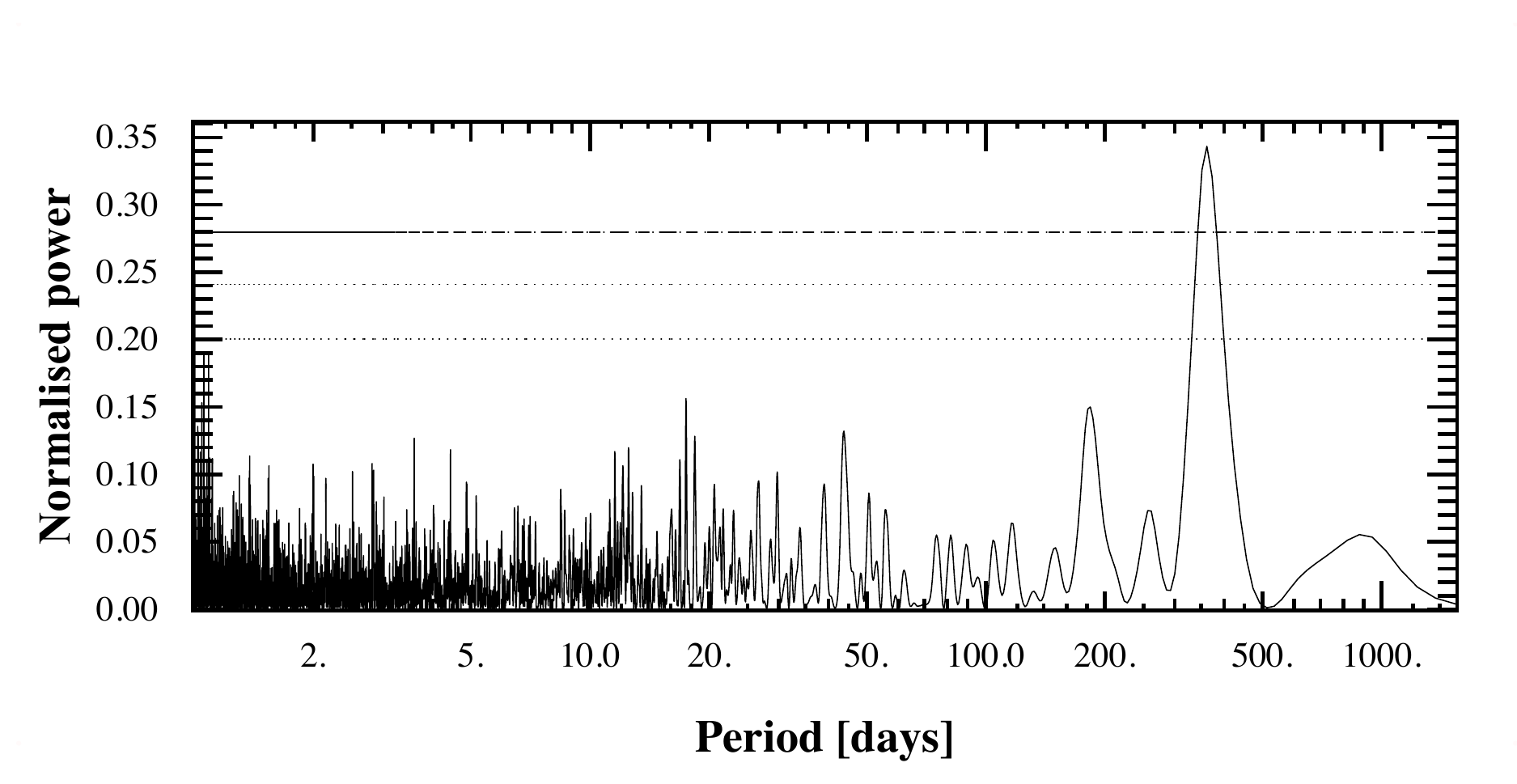}}
      \caption{WASP-41 Lomb-Scargle periodogram of the FWHM after subtracting the third-order trend. A clear peak is present at 1 year. The dotted lines correspond to a false-alarm probability of 0.1\%, 1\%, and 10\%.}
         \label{per_fw}
   \end{figure}

\begin{figure}
   \centering
   \resizebox{\hsize}{!}{\includegraphics{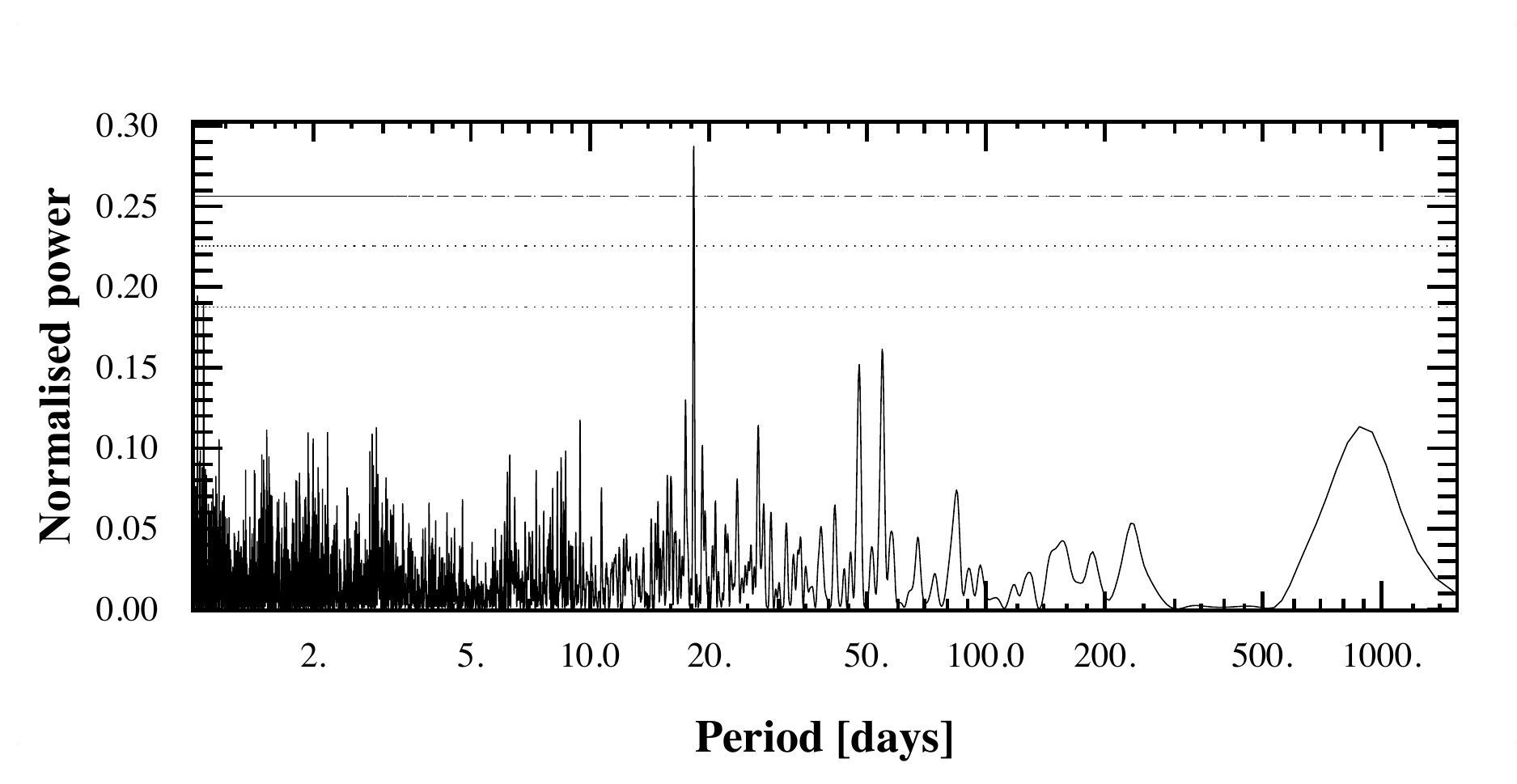}}
      \caption{WASP-41 Lomb-Scargle periodogram of the H$\alpha$ index after subtracting the third-order trend. A peak is present at $\sim$18 days, corresponding to the photometric rotation period. The dotted lines correspond to a false-alarm probability of 0.1\%, 1\%, and 10\%.}
         \label{per_ha}
   \end{figure}

The H$\alpha$ index reveals a time variation similar to the one observed on the FWHM (Fig. \ref{act_w41}) that can be modelled by a third-order trend in time. It is then reasonable to conclude that we are observing the magnetic cycle of WASP-41. After subtracting the third-order trend, we analysed the Lomb-Scargle periodogram of the residuals in the FWHM. A clear peak appears at exactly one year (see Fig. \ref{per_fw}). The same one-year signal in the FWHM has been observed on the brighter stars also regularly observed with CORALIE, as well as with similar instruments (SOPHIE, HARPS).
When we look at the Lomb-Scargle periodogram of the H$\alpha$ index after subtracting the long-term trend, a peak appears at $\sim$18 days (see Fig. \ref{per_ha}). Interestingly, this corresponds to the rotation period of the star derived from the WASP photometry by \citet{Maxted:2011fk}.

\begin{figure}
   \centering
      \includegraphics[width=8cm]{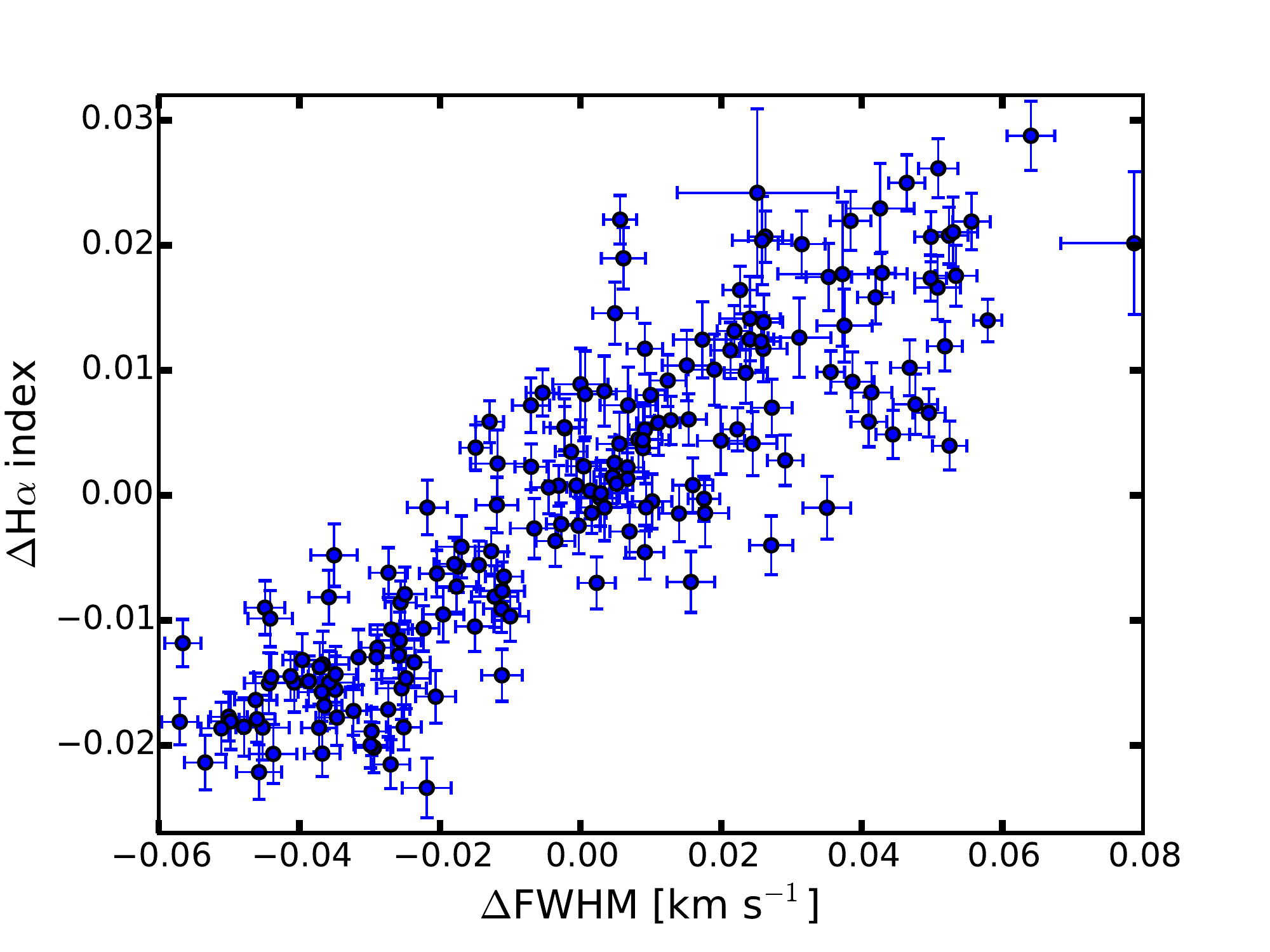}
      \caption{Correlation between the H$\alpha$ index and the FWHM from HARPS data of CoRoT-7.}
         \label{cor_c7}
   \end{figure}

Encouraged by the interesting result obtained using the H$\alpha$ index on WASP-41, we looked at the similarly active star \object{CoRoT-7} (with $\log R'_{\rm HK} = -4.62 \geq -4.7$). \citet{Queloz:2009yq} note that CoRoT-7 shows strong variability in the FWHM, comparable to what is observed on WASP-41. CoRoT-7 benefitted from simultaneous photometric and spectroscopic observations \citep[see][]{Haywood:2014lr}. We extracted the H$\alpha$ index from the HARPS spectra of CoRoT-7 in the same way as WASP-41. A very clear correlation is observed between the H$\alpha$ index and the FWHM (see Fig. \ref{cor_c7}). After subtracting a third-order trend, very likely due to the magnetic cycle, a signal appears at $\sim$23 days, corresponding to the photometric rotation period of the star. This similar result confirms that the H$\alpha$ index is a good activity indicator for such active stars.

The 400-day signal detected in the radial velocities is detected neither in  the FWHM nor in the H$\alpha$. Its origin is most likely due to the presence of a second planetary companion.

\subsection{Analysis}
As we did with WASP-47, we simultaneously fitted the transit light curves and the radial velocity measurements with an MCMC algorithm. To define the priors for the limb-darkening coefficients of the non-standard TRAPPIST $I+z$ filter, we took the average of the values of the standard filters $Ic$ and $z'$, and the quadratic sums of their errors. We did not include the FTS photometry from the discovery paper in our analysis because it suffered from poor weather conditions. As suggested by \citet{Gillon:2012lr}, we included a quadratic polynomial in time to model the transit light curves (except the fourth TRAPPIST and the first Danish transit observations). 

The observations with TRAPPIST on 21 Mar 2011 and 9 Mar 2012 suffered from malfunctions of the autofocus. As a consequence, the FWHM of the images varied a lot. To correct for this effect, we added a third-order and a quadratic polynomial in FWHM, respectively, to model those two light curves. The TRAPPIST observation on 19 Apr 2013 experienced a meridian flip that we modelled including an offset for that light curve. The TRAPPIST light curve from the 2 Apr 2011 and the Danish one from the 25 Apr 2013 obviously exhibited stellar spot crossings. We visually identified the affected parts of the light curves and decreased their weights by increasing the corresponding errors. In general, we applied a correction factor to the error bars of the light curves to account for red noise (values listed in Table. \ref{trans}). 

To improve the fit of the radial velocities and decrease the effects of stellar activity, we included a quadratic polynomial in the FWHM of the radial velocities from CORALIE. Then we quadratically added a "jitter" noise of 13.2\,m\,s$^{-1}$ to the error bars to equal the mean error with the standard deviation of the
best-fit model residuals. For the HARPS data (not including the RM sequence), we added a linear dependency in FWHM and a quadratic polynomial in bisector. We also added a "jitter" noise of 6.7\,m\,s$^{-1}$ to the radial velocity error bars. The baselines in FWHM and bisector were determined by comparing the BIC and the residual jitter obtained when running short chains with various models.

We modelled the Rossiter--McLaughlin effect following the equations of \citet{Gimenez:2006lr}. We used $\sqrt{v\,\sin I_{\star}}\,\cos\beta$, and $\sqrt{v\,\sin I_{\star}}\,\sin\beta$ as jump parameters, where $\beta$ is the projected angle between the stellar spin axis and the planet's orbital axis, and $v\,\sin I_{\star}$ is the projected rotational velocity. The Rossiter--McLaughlin sequence is fitted as an independent dataset to mitigate the effect of stellar activity. We did not need to add any "jitter" or dependency on other parameters to these data. Owing to the low transit impact parameter $b_{\rm tr}$, there was a strong degeneracy between $\beta$ and $v\,\sin I_{\star}$ \citep[see][]{Triaud:2011fk, Anderson:2011lr}. We imposed a normal prior on $v\,\sin\,I_{\star}$ to avoid unphysical solutions. Indeed from an MCMC with no prior on the $v\,\sin I_{\star}$ we obtain $\beta=-48\degr\pm29\degr$, and $v\,\sin\,I_{\star}=3.5^{+10.0}_{-1.0}$\,km\,s$^{-1}$ with $v\,\sin\,I_{\star}$ getting higher than 100\,km\,s$^{-1}$, clearly inconsistent with the spectral analysis. The value derived for $\beta$ is strongly dependent on the value of the prior for $v\,\sin\,I_{\star}$

The priors used for $T_{\rm eff}$ and [Fe/H] are defined with the values from \citet{Mortier:2013yq}. For the $v\,\sin I_{\star}$ we revised the value derived by \citet{Doyle:2014fk}, using the stellar parameters from \citet{Mortier:2013yq} to estimate the macroturbulence. Therefore we used $v\,\sin I_{\star}=2.66\pm0.28$\,km\,s$^{-1}$ as a normal prior. Since we benefitted from transit observations that span more than three years, we did not use a prior on the period and time of inferior conjunction $T_{\rm 0,b}$ of the transiting planet.

\subsection{Results}
\begin{figure}
   \centering
   \includegraphics[width=10cm]{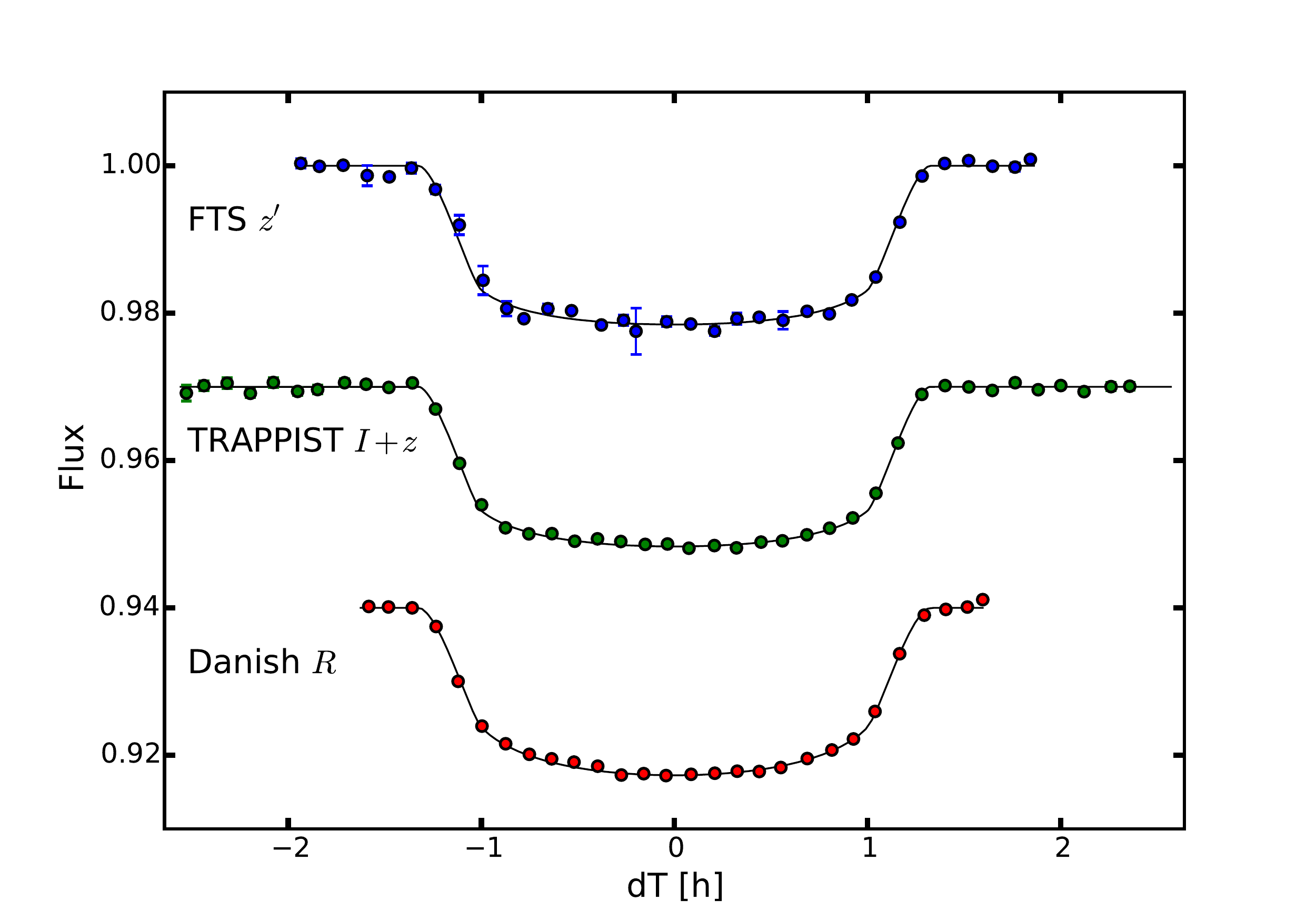}
      \caption{\object{WASP-41b} transit data (blue: FTS $z'$; green: TRAPPIST $I+z$; red: Danish $R$) and best fit model (black line). The TRAPPIST light curve was obtained by binning the data from five transit observations, and the Danish one from one full and one partial transit observations. Each point corresponds to a 7.2\,min bin.}
         \label{plot_photw41}
   \end{figure}

\begin{figure}
   \centering
   \includegraphics[width=10cm]{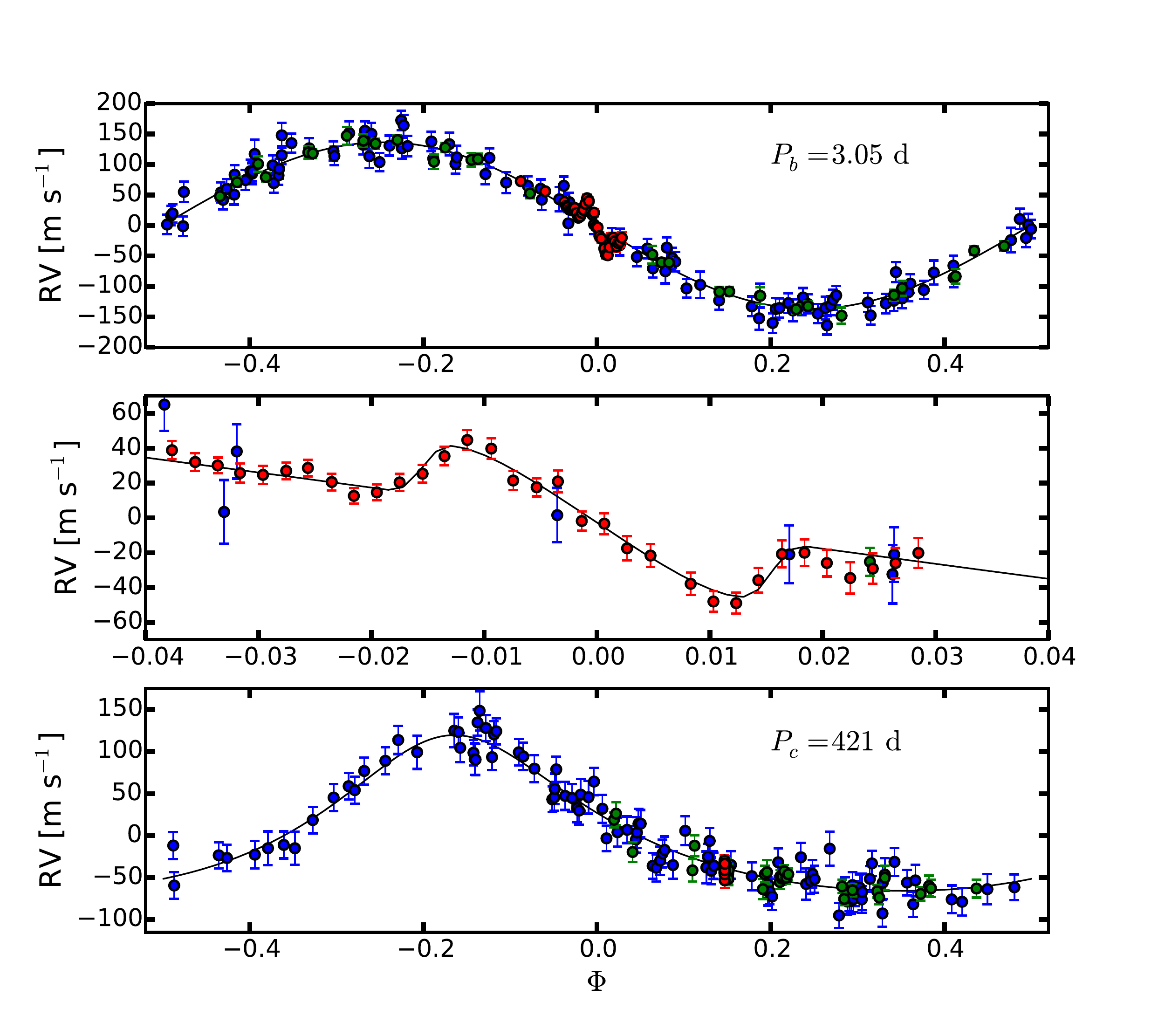}
      \caption{WASP-41 radial velocity data (blue: CORALIE, green: HARPS; red: HARPS RM sequence) and best fit model (black line). Top: Phase-folded on the period of the inner planet (outer planet subtracted). Middle: Zoom on the Rossiter--McLaughlin effect. Bottom: Phase-folded on the period of the outer planet (inner planet subtracted).}
         \label{plot_rvw41}
   \end{figure}

We present in Table \ref{res_w41} the median and 1 $\sigma$ limits of the marginalised posterior distributions after running five chains of 100\,000 accepted steps (considering a 20\% burn-in phase). The acceptance rate is $\sim$25\% ($\sim$40\% for the burn-in phase). The convergence of the chains is validated by the test of \citet{Gelman92} (PSFR better than 1.01), and the autocorrelations are shorter than 2\,000 steps. Including the second planet in the model undoubtedly improves the fit. The r.m.s. of the CORALIE data after the two-planet fit is ten times smaller than after a one-planet fit. The best fit solution is plotted in Fig. \ref{plot_photw41} for the transit light curves and in Fig. \ref{plot_rvw41} for the radial velocities. Again, the values provided in Table \ref{res_w41} are not the ones used in Figs. \ref{plot_rvw41} and \ref{plot_photw41}. We checked that they represent a self-consistent set of parameters by comparing the BIC using both sets of values (best fit and median of the posterior distribution) and get compatible results.

\begin{table}
\centering
\caption{Median and 1 $\sigma$ limits of the posterior marginalised PDFs
obtained for the WASP-41 system derived from our global MCMC
analysis}
\small
\begin{tabular}{lc}
\hline\hline
\multicolumn{2}{l}{\bf MCMC Jump parameters}\\
\hline
\multicolumn{2}{l}{\bf Star}\\
$T_{\rm eff}$ [K] & 5545 $\pm$ 33\\
{[Fe/H]} & 0.06 $\pm$ 0.02\\
$c_{1,z'}$ & 0.819 $\pm$ 0.009\\
$c_{2,z'}$ & $-0.230 \pm$ 0.006\\
$c_{1,I+z}$ & 0.85 $\pm$ 0.04\\
$c_{2,I+z}$ & $-0.19 \pm$ 0.04\\
$c_{1,R}$ & 1.08 $\pm$ 0.01\\
$c_{2,R}$ & $-0.09 \pm$ 0.01\\
\multicolumn{2}{l}{\bf Planet b}\\
Planet/star area ratio $(R_{\rm p}/R_{\star})^{2}$ [\%] & 1.87 $\pm$ 0.01\\
$b'=a_{\rm b} \cos i_{\rm p,b}/R_{\star}$ & 0.10 $\pm$ 0.06\\
$T_{\rm 14}$[d] & 0.1099$\pm$0.0002\\
$P_{\rm b}$ [d] & 3.0524040 $\pm$ 0.0000009\\
$T_{\rm 0,b}$ -- 2\,450\,000 [BJD$_{\rm TDB}$] & 6014.9936 $\pm$ 0.0001\\$\sqrt{e_{\rm b}}\,\cos\omega_{\rm b}$  & $0.002 \pm$ 0.048\\
$\sqrt{e_{\rm b}}\,\sin\omega_{\rm b}$ & 0.059 $\pm$ 0.074\\
$K_{\rm 2,b}=K_{\rm b}\sqrt{1-e_{\rm b}^{2}}\,P_{\rm b}^{1/3}$  [m s$^{-1}$ d$^{1/3}$] & 199.7 $\pm$ 2.3\\
$\sqrt{v\,\sin I_{\star}}\,\cos\beta$  & 1.42 $\pm$ 0.10\\
$\sqrt{v\,\sin I_{\star}}\,\sin\beta$ & $-0.78^{+0.38}_{-0.28}$\\
\multicolumn{2}{l}{\bf Planet c}\\
$P_{\rm c}$ [d] & 421 $\pm$ 2\\
$T_{\rm 0,c}$ -- 2\,450\,000 [BJD$_{\rm TDB}$] & 6011 $\pm$ 3\\
$\sqrt{\rm e_{c}}\,\cos\omega_{\rm c}$  & 0.53 $\pm$ 0.02\\
$\sqrt{\rm e_{c}}\,\sin\omega_{\rm c}$ & $-0.07 \pm$ 0.06\\
$K_{\rm 2,c}=K_{\rm c}\sqrt{1-e_{\rm c}^{2}}\,P_{\rm c}^{1/3}$ [m s$^{-1}$ d$^{1/3}$] & 676 $\pm$ 21\\
\hline
\multicolumn{2}{l}{\bf Derived parameters}\\
\hline
\multicolumn{2}{l}{\bf Star}\\
$u_{1,z'}$ & 0.282 $\pm$ 0.005\\
$u_{2,z'}$ & 0.256 $\pm$ 0.002\\
$u_{1,I+z}$ & 0.30 $\pm$ 0.02\\
$u_{2,I+z}$ & 0.25 $\pm$ 0.02\\
$u_{1,R}$ & 0.416 $\pm$ 0.006\\
$u_{2,R}$ & 0.252 $\pm$ 0.004\\
Density $\rho_{\star}$ [$\rho_{\sun}$] & 1.41 $\pm$ 0.05\\
Surface gravity $\log g_{\star}$ [cgs] & 4.53 $\pm$ 0.02\\
Mass $M_{\star}$ [$M_{\sun}$] & 0.93 $\pm$ 0.07\\
Radius $R_{\star}$ [$R_{\sun}$] & 0.87 $\pm$ 0.03\\
Luminosity $L_{\star}$ [$L_{\sun}$] & 0.64 $\pm$ 0.04\\
Projected rotation velocity $v\,\sin\,I_{\star}$ [km s$^{-1}$]\kern-20pt & 2.64 $\pm$ 0.25\\
\multicolumn{2}{l}{\bf Planet b}\\
Mass $M_{\rm p,b}$ [M$_{\rm Jup}$] & 0.94$\pm$0.05\\
Radius $R_{\rm p,b}$ [R$_{\rm Jup}$] & 1.18 $\pm$ 0.03\\
RV amplitude $K_{\rm b}$ [m\,s$^{-1}$] &138$\pm$2\\
Orbital eccentricity $e_{\rm b}$ & < 0.026 (at 2\,$\sigma$)\\
Orbital semi-major axis $a_{\rm b}$ [au]& 0.040 $\pm$ 0.001\\
Orbital inclination $i_{\rm p,b}$ [\degr]& 89.4 $\pm$ 0.3\\
Transit impact parameter $b_{\rm tr}$ & 0.10 $\pm$ 0.06\\
Projected orbital obliquity $\beta$ [\degr] & $-29^{+14}_{-10}$\\
Density $\rho_{\rm p,b}$ [$\rho_{\rm J}$] & 0.56 $\pm$ 0.03\\
Surface gravity $\log g_{\rm p,b}$ [cgs] & 3.24 $\pm$ 0.01\\
Equilibrium temperature $T_{\rm eq,b}^{(*)}$ [K] & 1244 $\pm$ 10\\
\multicolumn{2}{l}{\bf Planet c}\\
Minimum mass $M_{\rm p,c}\sin i_{\rm p,c}$ [M$_{\rm Jup}$] & 3.18 $\pm$ 0.20\\
RV amplitude $K_{\rm c}$ [m s$^{-1}$] & 94 $\pm$ 3\\
Orbital eccentricity $e_{\rm c}$ & 0.294 $\pm$ 0.024\\
Argument of periastron $\omega_{\rm c}$ [\degr] & 353 $\pm$ 6\\
Orbital semi-major axis $a_{\rm c}$ [au]& 1.07 $\pm$ 0.03\\
Equilibrium temperature $T_{\rm eq,b}^{(*)}$ [K] & 241 $\pm$ 2\\
$a_{\rm c}/R_{\star}$ & 265 $\pm$ 3\\
\hline
\end{tabular}
\label{res_w41}
\tablefoot{$^{(*)}$Assuming zero albedo and efficient redistribution of energy.}
\end{table}

\begin{figure}
   \centering
   \includegraphics[width=10cm]{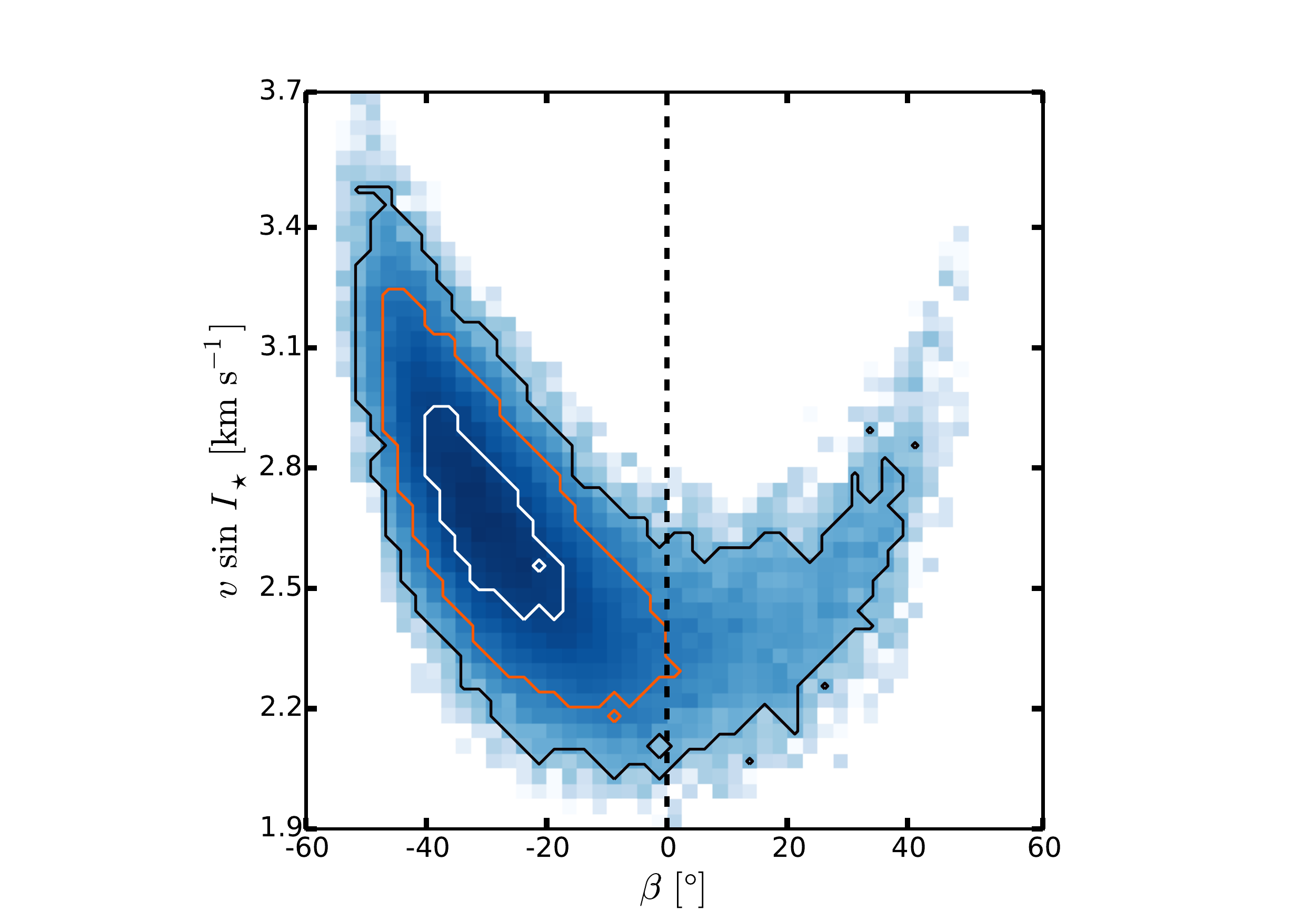}
      \caption{Posterior distribution of $\beta$ and $v\,\sin I_{\star}$ for WASP-41 b on logarithmic scale. We impose a prior of $2.66 \pm 0.28$\,km\,s$^{-1}$ on $v\,\sin I_{\star}$. The contours correspond to the 1-, 2-, and 3-$\sigma$ confidence limits for each parameter, and include 39.3\%, 86.5\%, and 98.9\% of the MCMC steps, respectively.}
         \label{rvsini_beta}
   \end{figure}

The best-fitting model suggests a misaligned planet with a projected orbital obliquity of  $\beta=-28\degr\pm13\degr$. However, the uncertainties are large (see Fig. \ref{rvsini_beta}), and the data do not allow an aligned orbit to be excluded ($P(|\beta|<20\degr)\sim24\%$). The rotation velocity expected from the photometric rotation period of the star ($\sim$ 2.44\,km\,s$^{-1}$) is compatible within the error bars with the spectroscopic $v\,\sin I_{\star}$ (2.64 $\pm$ 0.25\,km\,s$^{-1}$), suggesting that the star is equator-on. That reinforces the possibility of an aligned planet.

\section{Discussion} \label{disc}
{We have shown that finding long-period planets with instruments such as CORALIE is still possible around active stars. Recognising the periodic structure of activity-related signals is essential to avoiding false detections. While spots or plages induce signals at the rotation period of the star, the evolution of their coverage of the stellar surface is correlated with the magnetic cycle of the star. Such cycles have periods of several years and must be distinguished from long-period companions. \citet{Lovis:2011lr} studied the effect of magnetic cycles on the radial velocity measurements from the HARPS planet search sample. They derived empirical models of correlations between $R'_{\rm HK}$ and radial velocities, FWHM, contrast, and bisector spans, depending on the effective temperature $T_{\rm eff}$ and metallicity [Fe/H] of the star. However, their sample only contains 'quiet' stars because they are the preferred targets for planet searches. They consider only a handful of targets with a mean $\log R'_{\rm HK} > -4.7$, so their results cannot be applied to active stars. The CORALIE search for planets observes stars with a broader range of $\log R'_{\rm HK}$ than the HARPS one. While the emission in the Ca II (H \& K) lines is hard to extract from the CORALIE spectra, we have shown that the emission in the H$\alpha$, which is easy to measure, is a good indicator of activity for stars with $\log R'_{\rm HK} > -4.7$. A systematic study of the correlation between the H$\alpha$ emission and the radial velocities of active stars could help to estimate the influence of magnetic cycles on radial velocity measurements of active stars.

The transit technique has led to the discovery of more than 200 hot Jupiters, while only $\sim$30 have been discovered by Doppler surveys. A total of nine multiple planetary systems that include a hot Jupiter (with a < 0.1\,au) and a second planet with a fully probed orbit, are known (source: exoplanet.eu). Among them, four were discovered by radial velocity surveys and five by transit surveys. Those multiple systems represent $\sim$13\% of the known hot Jupiters found by radial velocities and only $\sim$2\% of those found by transit searches. The most obvious explanation for this difference is historical. Radial velocity surveys started discovering hot Jupiters well before transit surveys. The hot Jupiters discovered by radial velocities have been followed for more than ten years, allowing the detection of very wide companions (P > 9\,years). The dedicated large transit surveys (mainly HATNet and SuperWASP) started to efficiently discover hot Jupiters about six years ago, allowing only the characterisation of relatively close companions. If we exclusively consider companions within 2\,au, the rate of hot Jupiters with companions from the Doppler sample is divided by two.

The difference in the rates of detected additional planets in systems with a hot Jupiter may also be explained by the different observing strategies used in Doppler and transit surveys. The discovery of a hot Jupiter based on radial velocity data requires a large number of observations because the observer has no prior information on the planet. About typically ten radial velocity observations are carried out to confirm the existence
of a hot Jupiter previously detected by transit. In that case the observer knows the period of the planet and can plan the radial velocity observations in order to measure the radial velocity orbit very efficiently. This means that, on average, hot Jupiters discovered by the Doppler technique benefit from a larger number of radial velocity measurements than those discovered by transits. Increasing the number of radial velocity measurements makes the detection of longer period planetary companions easier.

In the beginning of the WASP follow-up with CORALIE, we started a systematic search for long-period companions around the WASP targets with a confirmed hot Jupiter. We have followed more than 100 WASP host stars for durations of two to eight years, including about 90 targets followed for more than three years. The publication of this work is in preparation (Neveu-VanMalle et al. in prep). In this sample only two stars with a hot Jupiter have been found to host a second planet that has completed at least one orbit. We still have a lack of multiple systems within 2\,au compared to the hot Jupiters discovered by radial velocity surveys. 

Detectability thresholds can contribute to explain the observed discrepancy. Figure \ref{mult_sys} shows the multiple planetary systems (with fully characterised orbits) that include a hot Jupiter, distributed in a plot of semi-major axis versus mass. The symbols located on the left-hand side of the plot (a <  0.1\,au) are the hot Jupiters, and the symbols on the right-hand side (a > 0.1\,AU) are the long-period planets. The dotted lines were drawn to visualise which planets are part of the same system. We notice that except for WASP-47 c, only the long-period planets from the Doppler-survey systems have semi-amplitudes smaller than 80\,m s$^{-1}$. Indeed, Doppler surveys target small samples of bright stars, while transit surveys target large samples of fainter stars. Besides being brighter, radial velocity targets are in general less active than transit targets. This means that a better precision can be obtained on the radial velocity measurements of stars from Doppler surveys. With a higher signal-to-noise ratio, detecting smaller signals is easier around brighter quiet stars. Smaller long-period planets might exist around stars with a known transiting hot Jupiter, but they are out of reach of current surveys.

  \begin{figure}
   \centering
   \includegraphics[width=10cm]{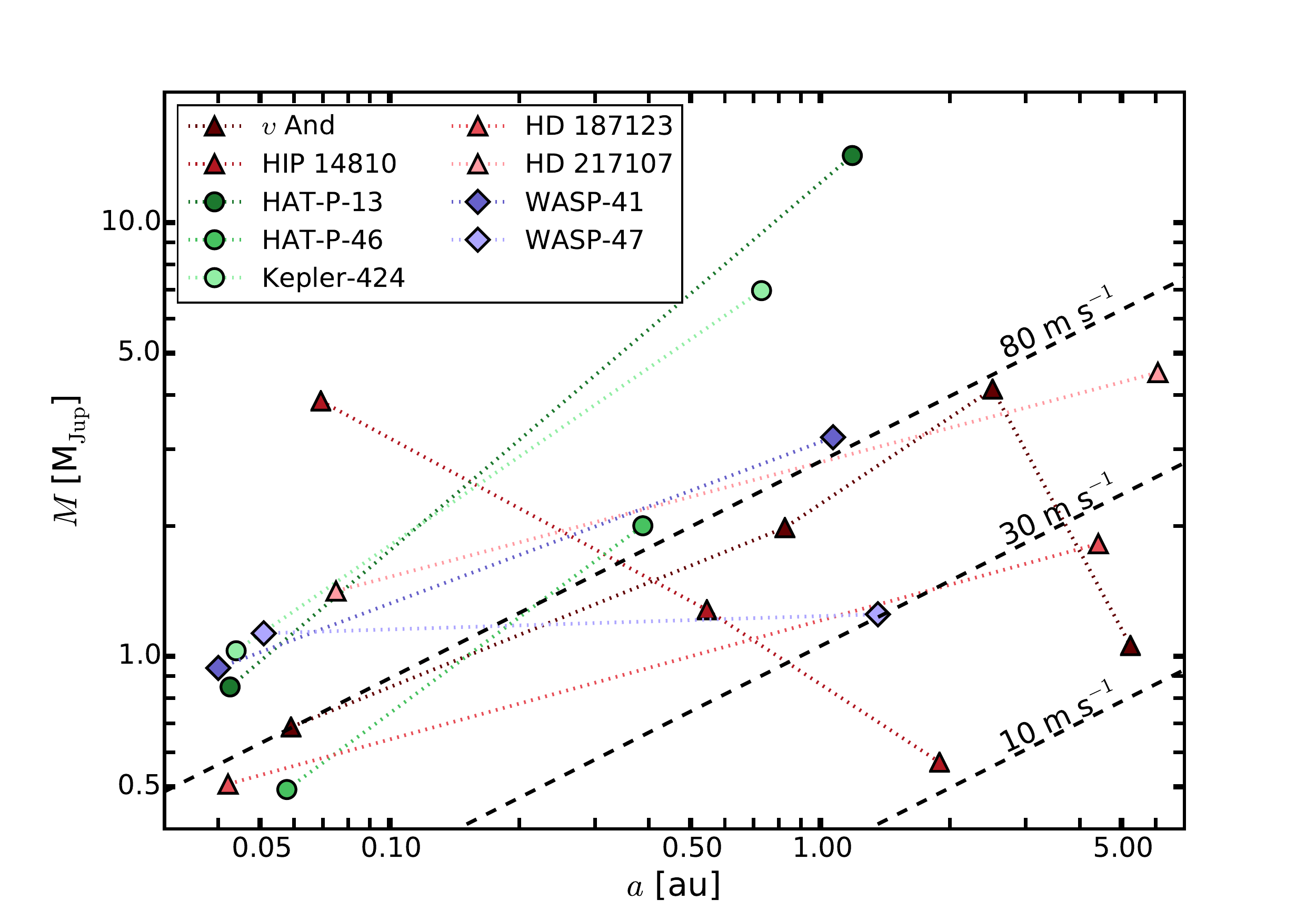}
      \caption{Known multiple planetary systems with fully probed orbits including a hot Jupiter (a < 0.1\,au). Triangles: hot Jupiter found by radial velocity surveys. Circles: Hot Jupiters found by transit surveys. Diamonds: systems presented in this paper. The black dashed lines correspond to a radial velocity semi-amplitude of 10\,m s$^{-1}$, 30\,m s$^{-1}$, and 80\,m s$^{-1}$ assuming a circular orbit around a 1-M$_{\sun}$ star. Most of these planets were detected only by radial velocities, thus we are plotting the minimum mass $M\sin i$ instead of the true mass $M$.}
         \label{mult_sys}
   \end{figure}
   
Knowing that WASP-47b and WASP-41b are transiting, it is natural to wonder if their outer planetary companions transit as well. If we assume that the inclination of the outer planet is uniformly distributed, independently of the inclination of the hot Jupiter, the transit probability is $\sim$0.4\% for WASP-47c and $\sim$0.6\% for WASP-41c \citep[using][]{Kane:2008lr}. If we assume a coplanar system where the orbit of the outer planet can be uniformly distributed around $5\degr$ from the orbit of the hot Jupiter, we obtain a transit probability of $\sim$6\% in both cases. Detecting the transit of the outer planet would bring strong evidence that the hot Jupiter had migrated inside the disk via gap opening \citep{Lin:1996lr,Marzari:2009lr}. However, a non-detection would not be sufficient to infer a high mutual inclination between the orbits of the planets, which would be evidence of a migration scenario involving dynamical interactions \citep{Rasio:1996fk,Fabrycky:2007fk,Nagasawa:2008uq,Matsumura:2010kx,Naoz:2011fj}.

The next predicted inferior conjunctions for WASP-47c are 2 Nov 2016 $\pm$ 30\,d and 24 May 2018 $\pm$ 43\,d, with an expected transit duration of $\sim$18\,h. A radial velocity follow-up is ongoing in order to improve the ephemeris. The next predicted inferior conjunction for WASP-41c (02 Nov 2016 $\pm$ 7\,d) cannot be observed by ground-based facilities because the star is close to the Sun. The two following ones are 12 Jan 2018 $\pm$ 8\,d and 3 Mar 2019 $\pm$ 10\,d with an expected transit duration of $\sim$13\,h.

Recently, \citep{Becker:2015lr} have announced the detection of two additional transiting planets around WASP-47. We checked {\bf if} the nine-day planet could be detected in our set of radial velocities. Unfortunately, the expected amplitude of the signal is below the sensitivity of CORALIE for such faint stars. The geometry of this system favours disc migration theories because scattering is unlikely to have kept the inner system coplanar. \citet{Sanchis-Ojeda:2015fk} have recently announced a low stellar obliquity for WASP-47 that reinforces the disc-migration hypothesis. Recent results have shown that Kozai interactions do not affect multi-planetary systems. In this case the outer planet could have `protected' the inner planets from secular interactions. This system brings fundamental constraints for planetary formation and migration theories.
 
}
  
\begin{acknowledgements}
M. Neveu-VanMalle thanks A. Doyle for her input. The {\it Euler} Swiss telescope is supported by the SNSF.
We thank the ESO staff at La Silla for their continuous help and support, along with the many observers on Euler and HARPS who collected the data used in that study. This work makes use of observations from the LCOGT network. TRAPPIST is a project funded by the Belgian Fund for Scientic Research (Fonds National de la Recherche Scientique, F.R.S.-FNRS) under grant FRFC 2.5.594.09.F, with the participation of the Swiss National Science Fundation (SNF). L. Delrez acknowledges support of the F.R.I.A. fund of the FNRS. M. Gillon and E. Jehin are FNRS Research Associates. ACC and CH acknowledge support from STFC grants ST/M001296/1 and ST/J001384/1 respectively. This work has made use of the Extrasolar Planet Encyclopaedia at exoplanet.eu \citep{Schneider:2011lr}.

\end{acknowledgements}


\bibliographystyle{aa}
\bibliography{biblio}

\begin{thebibliography}{45}
\expandafter\ifx\csname natexlab\endcsname\relax\def\natexlab#1{#1}\fi

\bibitem[{{Anderson} {et~al.}(2011){Anderson}, {Collier Cameron}, {Gillon},
  {Hellier}, {Jehin}, {Lendl}, {Queloz}, {Smalley}, {Triaud}, \&
  {Vanhuysse}}]{Anderson:2011lr}
{Anderson}, D.~R., {Collier Cameron}, A., {Gillon}, M., {et~al.} 2011, \aap,
  534, A16

\bibitem[{{Bakos} {et~al.}(2009){Bakos}, {Howard}, {Noyes}, {Hartman},
  {Torres}, {Kov{\'a}cs}, {Fischer}, {Latham}, {Johnson}, {Marcy}, {Sasselov},
  {Stefanik}, {Sip{\H o}cz}, {Kov{\'a}cs}, {Esquerdo}, {P{\'a}l},
  {L{\'a}z{\'a}r}, {Papp}, \& {S{\'a}ri}}]{Bakos:2009kx}
{Bakos}, G.~{\'A}., {Howard}, A.~W., {Noyes}, R.~W., {et~al.} 2009, \apj, 707,
  446

\bibitem[{{Becker} {et~al.}(2015){Becker}, {Vanderburg}, {Adams}, {Rappaport},
  \& {Schwengeler}}]{Becker:2015lr}
{Becker}, J.~C., {Vanderburg}, A., {Adams}, F.~C., {Rappaport}, S.~A., \&
  {Schwengeler}, H.~M. 2015, \apjl, 812, L18

\bibitem[{{Butler} {et~al.}(1997){Butler}, {Marcy}, {Williams}, {Hauser}, \&
  {Shirts}}]{Butler:1997fk}
{Butler}, R.~P., {Marcy}, G.~W., {Williams}, E., {Hauser}, H., \& {Shirts}, P.
  1997, \apjl, 474, L115

\bibitem[{{Cincunegui} {et~al.}(2007){Cincunegui}, {D{\'{\i}}az}, \&
  {Mauas}}]{Cincunegui:2007lr}
{Cincunegui}, C., {D{\'{\i}}az}, R.~F., \& {Mauas}, P.~J.~D. 2007, \aap, 469,
  309

\bibitem[{{Claret} \& {Bloemen}(2011)}]{Claret:2011qy}
{Claret}, A. \& {Bloemen}, S. 2011, \aap, 529, A75

\bibitem[{{Curiel} {et~al.}(2011){Curiel}, {Cant{\'o}}, {Georgiev},
  {Ch{\'a}vez}, \& {Poveda}}]{Curiel:2011qy}
{Curiel}, S., {Cant{\'o}}, J., {Georgiev}, L., {Ch{\'a}vez}, C.~E., \&
  {Poveda}, A. 2011, \aap, 525, A78

\bibitem[{{Doyle} {et~al.}(2014){Doyle}, {Davies}, {Smalley}, {Chaplin}, \&
  {Elsworth}}]{Doyle:2014fk}
{Doyle}, A.~P., {Davies}, G.~R., {Smalley}, B., {Chaplin}, W.~J., \&
  {Elsworth}, Y. 2014, \mnras, 444, 3592

\bibitem[{{Endl} {et~al.}(2014){Endl}, {Caldwell}, {Barclay}, {Huber},
  {Isaacson}, {Buchhave}, {Brugamyer}, {Robertson}, {Cochran}, {MacQueen},
  {Havel}, {Lucas}, {Howell}, {Fischer}, {Quintana}, \& {Ciardi}}]{Endl:2014vn}
{Endl}, M., {Caldwell}, D.~A., {Barclay}, T., {et~al.} 2014, \apj, 795, 151

\bibitem[{{Enoch} {et~al.}(2010){Enoch}, {Collier Cameron}, {Parley}, \&
  {Hebb}}]{Enoch:2010uq}
{Enoch}, B., {Collier Cameron}, A., {Parley}, N.~R., \& {Hebb}, L. 2010, \aap,
  516, A33

\bibitem[{{Fabrycky} \& {Tremaine}(2007)}]{Fabrycky:2007fk}
{Fabrycky}, D. \& {Tremaine}, S. 2007, \apj, 669, 1298

\bibitem[{{Feng} {et~al.}(2015){Feng}, {Wright}, {Nelson}, {Wang}, {Ford},
  {Marcy}, {Isaacson}, \& {Howard}}]{Feng2015}
{Feng}, Y.~K., {Wright}, J.~T., {Nelson}, B., {et~al.} 2015, \apj, 800, 22

\bibitem[{Gelman \& Rubin(1992)}]{Gelman92}
Gelman, A. \& Rubin, D. 1992, Statistical Science, 7, 457

\bibitem[{{Gillon} {et~al.}(2011){Gillon}, {Doyle}, {Lendl}, {Maxted},
  {Triaud}, {Anderson}, {Barros}, {Bento}, {Collier-Cameron}, {Enoch}, {Faedi},
  {Hellier}, {Jehin}, {Magain}, {Montalb{\'a}n}, {Pepe}, {Pollacco}, {Queloz},
  {Smalley}, {Segransan}, {Smith}, {Southworth}, {Udry}, {West}, \&
  {Wheatley}}]{Gillon:2011fj}
{Gillon}, M., {Doyle}, A.~P., {Lendl}, M., {et~al.} 2011, \aap, 533, A88

\bibitem[{{Gillon} {et~al.}(2012){Gillon}, {Triaud}, {Fortney}, {Demory},
  {Jehin}, {Lendl}, {Magain}, {Kabath}, {Queloz}, {Alonso}, {Anderson},
  {Collier Cameron}, {Fumel}, {Hebb}, {Hellier}, {Lanotte}, {Maxted},
  {Mowlavi}, \& {Smalley}}]{Gillon:2012lr}
{Gillon}, M., {Triaud}, A.~H.~M.~J., {Fortney}, J.~J., {et~al.} 2012, \aap,
  542, A4

\bibitem[{{Gim{\'e}nez}(2006)}]{Gimenez:2006lr}
{Gim{\'e}nez}, A. 2006, \apj, 650, 408

\bibitem[{{Gomes da Silva} {et~al.}(2014){Gomes da Silva}, {Santos}, {Boisse},
  {Dumusque}, \& {Lovis}}]{Gomes-da-Silva:2014qy}
{Gomes da Silva}, J., {Santos}, N.~C., {Boisse}, I., {Dumusque}, X., \&
  {Lovis}, C. 2014, \aap, 566, A66

\bibitem[{{Hartman} {et~al.}(2014){Hartman}, {Bakos}, {Torres}, {Kov{\'a}cs},
  {Johnson}, {Howard}, {Marcy}, {Latham}, {Bieryla}, {Buchhave}, {Bhatti},
  {B{\'e}ky}, {Csubry}, {Penev}, {de Val-Borro}, {Noyes}, {Fischer},
  {Esquerdo}, {Everett}, {Szklen{\'a}r}, {Zhou}, {Bayliss}, {Shporer},
  {Fulton}, {Sanchis-Ojeda}, {Falco}, {L{\'a}z{\'a}r}, {Papp}, \&
  {S{\'a}ri}}]{Hartman:2014yq}
{Hartman}, J.~D., {Bakos}, G.~{\'A}., {Torres}, G., {et~al.} 2014, \aj, 147,
  128

\bibitem[{{Haywood} {et~al.}(2014){Haywood}, {Collier Cameron}, {Queloz},
  {Barros}, {Deleuil}, {Fares}, {Gillon}, {Lanza}, {Lovis}, {Moutou}, {Pepe},
  {Pollacco}, {Santerne}, {S{\'e}gransan}, \& {Unruh}}]{Haywood:2014lr}
{Haywood}, R.~D., {Collier Cameron}, A., {Queloz}, D., {et~al.} 2014, \mnras,
  443, 2517

\bibitem[{{Hellier} {et~al.}(2012){Hellier}, {Anderson}, {Collier Cameron},
  {Doyle}, {Fumel}, {Gillon}, {Jehin}, {Lendl}, {Maxted}, {Pepe}, {Pollacco},
  {Queloz}, {S{\'e}gransan}, {Smalley}, {Smith}, {Southworth}, {Triaud},
  {Udry}, \& {West}}]{Hellier:2012uq}
{Hellier}, C., {Anderson}, D.~R., {Collier Cameron}, A., {et~al.} 2012, \mnras,
  426, 739

\bibitem[{{Holman} {et~al.}(2006){Holman}, {Winn}, {Latham}, {O'Donovan},
  {Charbonneau}, {Bakos}, {Esquerdo}, {Hergenrother}, {Everett}, \&
  {P{\'a}l}}]{Holman:2006fk}
{Holman}, M.~J., {Winn}, J.~N., {Latham}, D.~W., {et~al.} 2006, \apj, 652, 1715

\bibitem[{{Jehin} {et~al.}(2011){Jehin}, {Gillon}, {Queloz}, {Magain},
  {Manfroid}, {Chantry}, {Lendl}, {Hutsem{\'e}kers}, \& {Udry}}]{Jehin:2011fj}
{Jehin}, E., {Gillon}, M., {Queloz}, D., {et~al.} 2011, The Messenger, 145, 2

\bibitem[{{Kane} \& {von Braun}(2008)}]{Kane:2008lr}
{Kane}, S.~R. \& {von Braun}, K. 2008, \apj, 689, 492

\bibitem[{{Kipping}(2013)}]{Kipping2013}
{Kipping}, D.~M. 2013, \mnras, 435, 2152

\bibitem[{{Knutson} {et~al.}(2014){Knutson}, {Fulton}, {Montet}, {Kao}, {Ngo},
  {Howard}, {Crepp}, {Hinkley}, {Bakos}, {Batygin}, {Johnson}, {Morton}, \&
  {Muirhead}}]{Knutson:2014lr}
{Knutson}, H.~A., {Fulton}, B.~J., {Montet}, B.~T., {et~al.} 2014, \apj, 785,
  126

\bibitem[{{Lin} {et~al.}(1996){Lin}, {Bodenheimer}, \&
  {Richardson}}]{Lin:1996lr}
{Lin}, D.~N.~C., {Bodenheimer}, P., \& {Richardson}, D.~C. 1996, \nat, 380, 606

\bibitem[{{Lovis} {et~al.}(2011){Lovis}, {Dumusque}, {Santos}, {Bouchy},
  {Mayor}, {Pepe}, {Queloz}, {S{\'e}gransan}, \& {Udry}}]{Lovis:2011lr}
{Lovis}, C., {Dumusque}, X., {Santos}, N.~C., {et~al.} 2011, ArXiv e-prints:
  1107.5325

\bibitem[{{Marzari} \& {Nelson}(2009)}]{Marzari:2009lr}
{Marzari}, F. \& {Nelson}, A.~F. 2009, \apj, 705, 1575

\bibitem[{{Matsumura} {et~al.}(2010){Matsumura}, {Peale}, \&
  {Rasio}}]{Matsumura:2010kx}
{Matsumura}, S., {Peale}, S.~J., \& {Rasio}, F.~A. 2010, \apj, 725, 1995

\bibitem[{{Maxted} {et~al.}(2011){Maxted}, {Anderson}, {Collier Cameron},
  {Hellier}, {Queloz}, {Smalley}, {Street}, {Triaud}, {West}, {Gillon},
  {Lister}, {Pepe}, {Pollacco}, {S{\'e}gransan}, {Smith}, \&
  {Udry}}]{Maxted:2011fk}
{Maxted}, P.~F.~L., {Anderson}, D.~R., {Collier Cameron}, A., {et~al.} 2011,
  \pasp, 123, 547

\bibitem[{{Meunier} \& {Delfosse}(2009)}]{Meunier:2009fk}
{Meunier}, N. \& {Delfosse}, X. 2009, \aap, 501, 1103

\bibitem[{{Mortier} {et~al.}(2013){Mortier}, {Santos}, {Sousa}, {Fernandes},
  {Adibekyan}, {Delgado Mena}, {Montalto}, \& {Israelian}}]{Mortier:2013yq}
{Mortier}, A., {Santos}, N.~C., {Sousa}, S.~G., {et~al.} 2013, \aap, 558, A106

\bibitem[{{Nagasawa} {et~al.}(2008){Nagasawa}, {Ida}, \&
  {Bessho}}]{Nagasawa:2008uq}
{Nagasawa}, M., {Ida}, S., \& {Bessho}, T. 2008, \apj, 678, 498

\bibitem[{{Naoz} {et~al.}(2011){Naoz}, {Farr}, {Lithwick}, {Rasio}, \&
  {Teyssandier}}]{Naoz:2011fj}
{Naoz}, S., {Farr}, W.~M., {Lithwick}, Y., {Rasio}, F.~A., \& {Teyssandier}, J.
  2011, \nat, 473, 187

\bibitem[{{Pollacco} {et~al.}(2006){Pollacco}, {Skillen}, {Collier Cameron},
  {Christian}, {Hellier}, {Irwin}, {Lister}, {Street}, {West}, {Anderson},
  {Clarkson}, {Deeg}, {Enoch}, {Evans}, {Fitzsimmons}, {Haswell}, {Hodgkin},
  {Horne}, {Kane}, {Keenan}, {Maxted}, {Norton}, {Osborne}, {Parley}, {Ryans},
  {Smalley}, {Wheatley}, \& {Wilson}}]{Pollacco:2006qy}
{Pollacco}, D.~L., {Skillen}, I., {Collier Cameron}, A., {et~al.} 2006, \pasp,
  118, 1407

\bibitem[{{Queloz} {et~al.}(2010){Queloz}, {Anderson}, {Collier Cameron},
  {Gillon}, {Hebb}, {Hellier}, {Maxted}, {Pepe}, {Pollacco}, {S{\'e}gransan},
  {Smalley}, {Triaud}, {Udry}, \& {West}}]{Queloz:2010lr}
{Queloz}, D., {Anderson}, D.~R., {Collier Cameron}, A., {et~al.} 2010, \aap,
  517, L1

\bibitem[{{Queloz} {et~al.}(2009){Queloz}, {Bouchy}, {Moutou}, {Hatzes},
  {H{\'e}brard}, {Alonso}, {Auvergne}, {Baglin}, {Barbieri}, {Barge}, {Benz},
  {Bord{\'e}}, {Deeg}, {Deleuil}, {Dvorak}, {Erikson}, {Ferraz Mello},
  {Fridlund}, {Gandolfi}, {Gillon}, {Guenther}, {Guillot}, {Jorda}, {Hartmann},
  {Lammer}, {L{\'e}ger}, {Llebaria}, {Lovis}, {Magain}, {Mayor}, {Mazeh},
  {Ollivier}, {P{\"a}tzold}, {Pepe}, {Rauer}, {Rouan}, {Schneider},
  {Segransan}, {Udry}, \& {Wuchterl}}]{Queloz:2009yq}
{Queloz}, D., {Bouchy}, F., {Moutou}, C., {et~al.} 2009, \aap, 506, 303

\bibitem[{{Rasio} \& {Ford}(1996)}]{Rasio:1996fk}
{Rasio}, F.~A. \& {Ford}, E.~B. 1996, Science, 274, 954

\bibitem[{{Sanchis-Ojeda} {et~al.}(2015){Sanchis-Ojeda}, {Winn}, {Dai},
  {Howard}, {Isaacson}, {Marcy}, {Petigura}, {Sinukoff}, {Weiss}, {Albrecht},
  {Hirano}, \& {Rogers}}]{Sanchis-Ojeda:2015fk}
{Sanchis-Ojeda}, R., {Winn}, J.~N., {Dai}, F., {et~al.} 2015, \apjl, 812, L11

\bibitem[{{Schneider} {et~al.}(2011){Schneider}, {Dedieu}, {Le Sidaner},
  {Savalle}, \& {Zolotukhin}}]{Schneider:2011lr}
{Schneider}, J., {Dedieu}, C., {Le Sidaner}, P., {Savalle}, R., \&
  {Zolotukhin}, I. 2011, \aap, 532, A79

\bibitem[{{Southworth}(2011)}]{Southworth:2011kx}
{Southworth}, J. 2011, \mnras, 417, 2166

\bibitem[{{Triaud} {et~al.}(2011){Triaud}, {Queloz}, {Hellier}, {Gillon},
  {Smalley}, {Hebb}, {Collier Cameron}, {Anderson}, {Boisse}, {H{\'e}brard},
  {Jehin}, {Lister}, {Lovis}, {Maxted}, {Pepe}, {Pollacco}, {S{\'e}gransan},
  {Simpson}, {Udry}, \& {West}}]{Triaud:2011fk}
{Triaud}, A.~H.~M.~J., {Queloz}, D., {Hellier}, C., {et~al.} 2011, \aap, 531,
  A24

\bibitem[{{Winn} {et~al.}(2010){Winn}, {Fabrycky}, {Albrecht}, \&
  {Johnson}}]{Winn:2010lr}
{Winn}, J.~N., {Fabrycky}, D., {Albrecht}, S., \& {Johnson}, J.~A. 2010, \apjl,
  718, L145

\bibitem[{{Wright} {et~al.}(2009){Wright}, {Fischer}, {Ford}, {Veras}, {Wang},
  {Henry}, {Marcy}, {Howard}, \& {Johnson}}]{Wright:2009fj}
{Wright}, J.~T., {Fischer}, D.~A., {Ford}, E.~B., {et~al.} 2009, \apjl, 699,
  L97

\bibitem[{{Wright} {et~al.}(2007){Wright}, {Marcy}, {Fischer}, {Butler},
  {Vogt}, {Tinney}, {Jones}, {Carter}, {Johnson}, {McCarthy}, \&
  {Apps}}]{Wright:2007uq}
{Wright}, J.~T., {Marcy}, G.~W., {Fischer}, D.~A., {et~al.} 2007, \apj, 657,
  533

\end{thebibliography}

\Online
\begin{appendix}

\section{Additional plots}
\onlfig{
\begin{figure*}
   \centering
   \includegraphics[width=\textwidth]{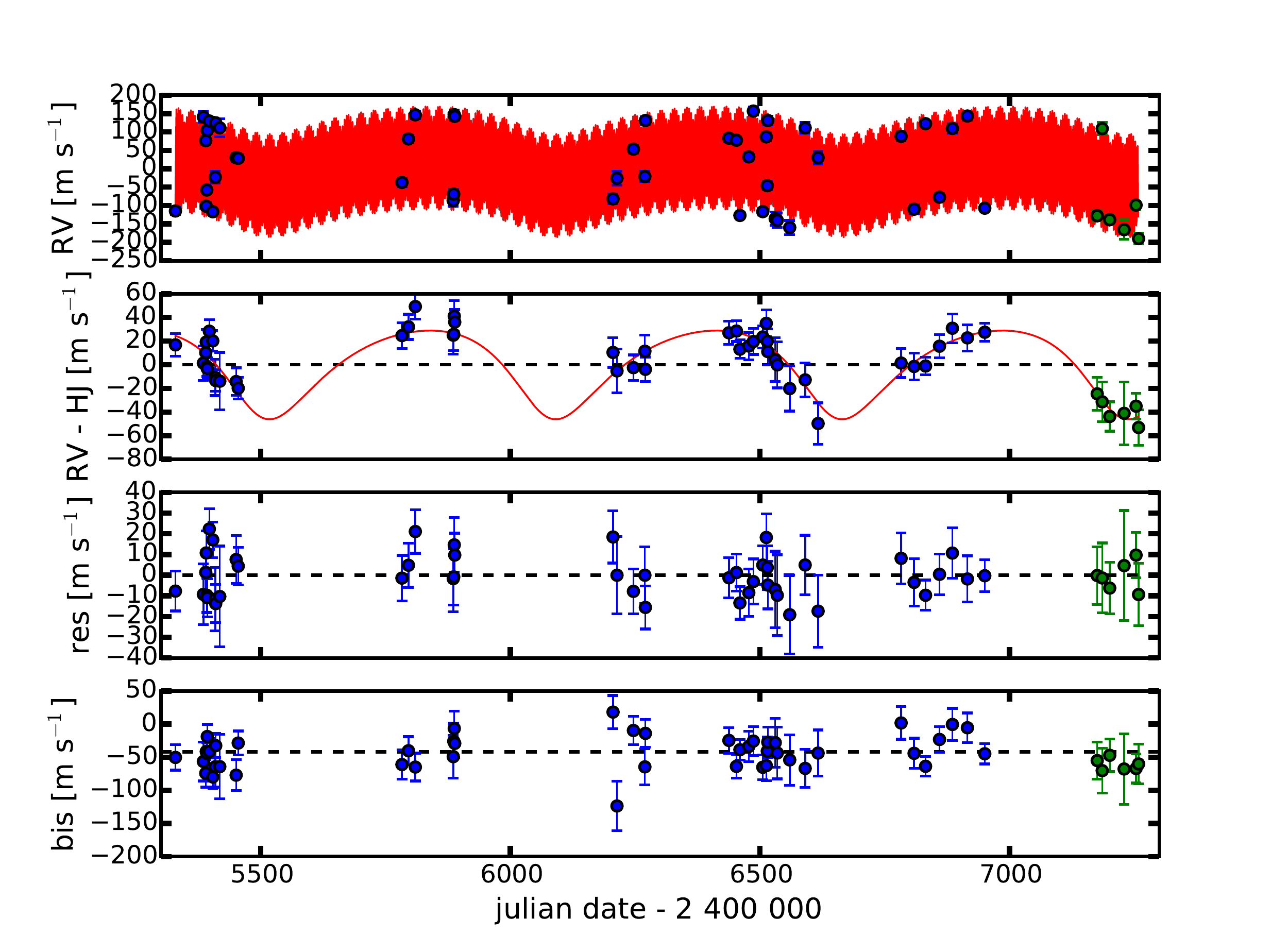}
      \caption{CORALIE data of WASP-47 plotted in time. Blue/Green dots: before/after the upgrade. Top panel: Radial velocities superimposed with the best fit model. Second panel: Residuals of the radial velocities after subtracting the hot Jupiter superimposed with the model for the second planet. Third panel: Residuals of the radial velocities after subtracting the two planets. Bottom panel: Bisector spans.}
         \label{w47_ap}
   \end{figure*}
   }
   \onlfig{
   \begin{figure*}
   \centering
   \includegraphics[width=\textwidth]{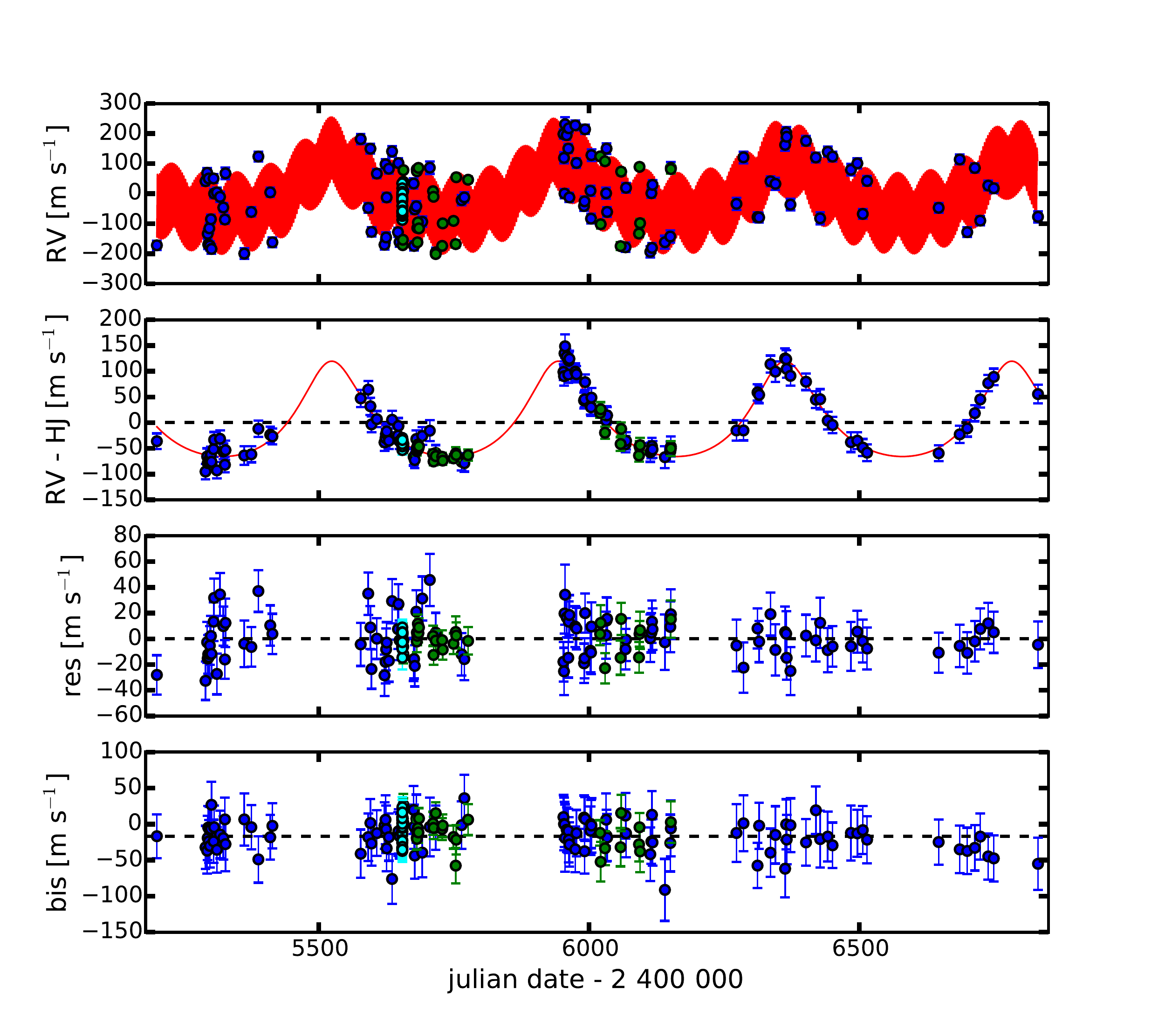}
      \caption{Radial velocity data of WASP-41 plotted in time. Blue: CORALIE; Green: HARPS; Cyan: RM sequence. Top panel: Radial velocities superimposed with the best fit model. Second panel: Residuals of the radial velocities after subtracting the hot Jupiter superimposed with the model for the second planet. Third panel: Residuals of the radial velocities after subtracting the two planets. Bottom panel: Bisector spans.}
         \label{w41_ap}
   \end{figure*}
   }
   \onlfig{
   \begin{figure}
   \centering
   \includegraphics[width=10cm]{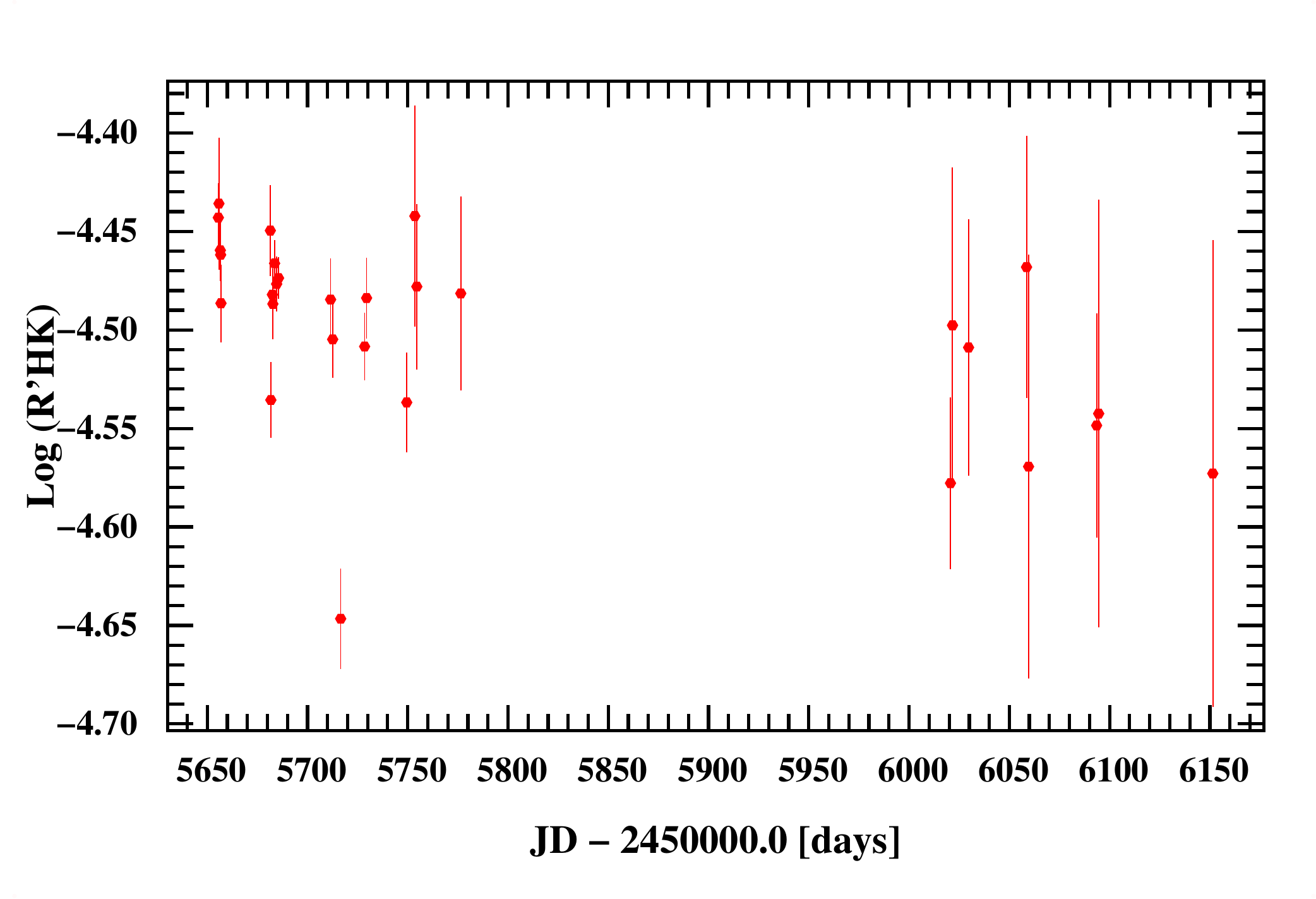}
      \caption{WASP-41 $\log R'_{\rm HK}$ extracted from the HARPS spectra.}
         \label{w41_rhk}
   \end{figure}
   }

\section{MCMC initial values and priors}

\onltab{
\begin{table*}
\centering
\caption{Initial values and priors for WASP-47}
\begin{tabular}{lcc}
\hline\hline
{\bf Parameter} & {\bf Value} & {\bf Prior }\\
\hline
$M_{\star}$ [$M_{\sun}$] & 1.07 $\pm$ 0.10 $^{1}$ & $\ge 0$ \\
$T_{\rm eff}$ [K] & 5576 $\pm$ 68 $^{1}$ & Gaussian\\
{[Fe/H]} & 0.36 $\pm$ 0.05 $^{1}$ & Gaussian\\
$u_{1}$ & 0.4584 $\pm$  0.0144 $^{2}$ & Gaussian, $u_{1} + u_{2}<1$  $^{4}$\\
$u_{2}$ & 0.2417 $\pm$ 0.0086 $^{2}$ & Gaussian,  $u_{1,} + u_{2}<1$  $^{4}$\\
$P_{\rm b}$ [d] & 4.1591399 $\pm$0.0000072 $^{3}$ & Gaussian\\
$T_{\rm 0,b}$ -- 2\,450\,000 [BJD$_{\rm TDB}$] & 5764.34602 $\pm$ 0.00022 $^{3}$ & Gaussian\\
$(R_{\rm p}/R_{\star})^{2}$ [\%] & 1.051 $\pm$ 0.014 $^{3}$ & uniform $\ge 0$\\
$T_{\rm 14}$[d] & 0.14933 $\pm$ 0.00065 $^{3}$ & uniform $\ge 0$\\
$b_{\rm tr}$ & 0.14 $\pm$ 0.11 $^{3}$ & uniform on $b'=a_{\rm b}\cos i_{\rm p,b}/R_{\star}$, $0 \le b_{\rm tr} \le a_{\rm b}/R_{\star}$\\
$K_{\rm b}$ [m\,s$^{-1}$] & 136.0 $\pm$ 5 $^{3}$ & uniform on $K_{\rm 2,b}=K_{\rm b}\sqrt{1-e_{\rm b}^{2}}\,P_{\rm b}^{1/3}$ and $\ge 0$\\
$e_{\rm b}$ & 0 & uniform on $\sqrt{e_{\rm b}}\,\cos\omega_{\rm b}$ and $\sqrt{e_{\rm b}}\,\sin\omega_{\rm b}$, if > 1 then $e_{\rm b}$ = 0.999\\
$\omega_{\rm b}$ & 0 & uniform on $\sqrt{e_{\rm b}}\,\cos\omega_{\rm b}$ and $\sqrt{e_{\rm b}}\,\sin\omega_{\rm b}$\\
$P_{\rm c}$ [d] & 571 $\pm$ 8 & uniform 0\,d $\le P \le$ 1000\,yr\\
$T_{\rm 0,c}$ -- 2\,450\,000 [BJD$_{\rm TDB}$] & 5983 $\pm$ 26 & uniform\\
$K_{\rm c}$ [m\,s$^{-1}$] & 34 $\pm$ 10 & uniform on $K_{\rm 2,c}=K_{\rm c}\sqrt{1-e_{\rm c}^{2}}\,P_{\rm c}^{1/3}$ and $\ge 0$\\
$e_{\rm c}$ & 0.2 $\pm$ 0.2 & uniform on $\sqrt{e_{\rm c}}\,\cos\omega_{\rm c}$ and $\sqrt{e_{\rm c}}\,\sin\omega_{\rm c}$, if > 1 then $e_{\rm c}$ = 0.999\\
$\omega_{\rm c}$ & 0 & uniform on $\sqrt{e_{\rm c}}\,\cos\omega_{\rm c}$ and $\sqrt{e_{\rm c}}\,\sin\omega_{\rm c}$\\
\hline
\end{tabular}
\tablebib{$^{ 1 }$ From \citet{Mortier:2013yq}. $^{ 2 }$ Interpolated from \cite{Claret:2011qy}. $^{ 3 }$ From \citet{Hellier:2012uq}.  $^{5}$ See \citet{Kipping2013}.}
\label{priors_w47}
\end{table*}
}

\onltab{
\begin{table*}
\centering
\caption{Initial values and priors for WASP-41}
\begin{tabular}{lcc}
\hline\hline
{\bf Parameter} & {\bf Value} & {\bf Prior }\\
\hline
$M_{\star}$ [$M_{\sun}$] & 0.90 $\pm$ 0.07 $^{1}$ & $\ge 0$ \\
$T_{\rm eff}$ [K] & 5546 $\pm$ 33 $^{1}$ & Gaussian\\
{[Fe/H]} & 0.06 $\pm$ 0.02 $^{1}$ & Gaussian\\
$u_{1,z'}$ & 0.2817 $\pm$  0.0044 $^{2}$ & Gaussian, $u_{1,z'} + u_{2,z'}<1$ $^{5}$\\
$u_{2,z'}$ & 0.2559 $\pm$  0.0022 $^{2}$ & Gaussian,  $u_{1,z'} + u_{2,z'}<1$ $^{5}$\\
$u_{1,I+z}$ & 0.3090 $\pm$  0.0340 $^{2}$ & Gaussian, $u_{1,I+z} + u_{2,I+z}<1$ $^{5}$\\
$u_{2,I+z}$ & 0.2510 $\pm$  0.0170 $^{2}$ & Gaussian, $u_{1,I+z} + u_{2,I+z}<1$ $^{5}$\\
$u_{1,R}$ & 0.4144 $\pm$  0.0065 $^{2}$ & Gaussian, $u_{1,R} + u_{2,R}<1$ $^{5}$\\
$u_{2,R}$ & 0.2520 $\pm$  0.0037 $^{2}$ & Gaussian, $u_{1,R} + u_{2,R}<1$ $^{5}$\\
$v\,\sin\,I_{\star}$ [km s$^{-1}$] & 2.66 $\pm$ 0.28 $^{3}$ & Gaussian\\
$P_{\rm b}$ [d] & 3.052401 $\pm$ 0.000004 $^{4}$ & uniform 0\,d $\le P \le$ 1000\,yr\\
$T_{\rm 0,b}$ -- 2\,450\,000 [BJD$_{\rm TDB}$] & 6014.991 $\pm$ 0.001 $^{4}$ & uniform\\
$(R_{\rm p}/R_{\star})^{2}$ [\%] & 1.86 $\pm$ 0.04 $^{4}$ & uniform $\ge 0$\\
$T_{\rm 14}$[d] & 0.108 $\pm$ 0.002 $^{4}$ & uniform $\ge 0$\\
$b_{\rm tr}$ & 0.40 $\pm$ 0.15 $^{4}$ & uniform on $b'=a_{\rm b}\cos i_{\rm p,b}/R_{\star}$, $0 \le b_{\rm tr} \le a_{\rm b}/R_{\star}$\\
$K_{\rm b}$ [m\,s$^{-1}$] & 135.0 $\pm$ 8 $^{4}$ & uniform on $K_{\rm 2,b}=K_{\rm b}\sqrt{1-e_{\rm b}^{2}}\,P_{\rm b}^{1/3}$ and $\ge 0$\\
$e_{\rm b}$ & 0 & uniform on $\sqrt{e_{\rm b}}\,\cos\omega_{\rm b}$ and $\sqrt{e_{\rm b}}\,\sin\omega_{\rm b}$, if > 1 then $e_{\rm b}$ = 0.999\\
$\omega_{\rm b}$ & 0 & uniform on $\sqrt{e_{\rm b}}\,\cos\omega_{\rm b}$ and $\sqrt{e_{\rm b}}\,\sin\omega_{\rm b}$\\
$\beta$ [\degr] & 0 & uniform on $\sqrt{v\,\sin I_{\star}}\,\cos\beta$ and $\sqrt{v\,\sin I_{\star}}\,\sin\beta$\\
$P_{\rm c}$ [d] & 418 $\pm$ 3 & uniform 0\,d $\le P \le$ 1000\,yr\\
$T_{\rm 0,c}$ -- 2\,450\,000 [BJD$_{\rm TDB}$] & 6006 $\pm$ 10 & uniform\\
$K_{\rm c}$ [m\,s$^{-1}$] & 94 $\pm$ 4 & uniform on $K_{\rm 2,c}=K_{\rm c}\sqrt{1-e_{\rm c}^{2}}\,P_{\rm c}^{1/3}$ and $\ge 0$\\
$e_{\rm c}$ & 0.3 $\pm$ 0.05 & uniform on $\sqrt{e_{\rm c}}\,\cos\omega_{\rm c}$ and $\sqrt{e_{\rm c}}\,\sin\omega_{\rm c}$, if > 1 then $e_{\rm c}$ = 0.999\\
$\omega_{\rm c}$ & 0 & uniform on $\sqrt{e_{\rm c}}\,\cos\omega_{\rm c}$ and $\sqrt{e_{\rm c}}\,\sin\omega_{\rm c}$\\
\hline
\end{tabular}
\tablebib{$^{ 1 }$ From \citet{Mortier:2013yq}. $^{ 2 }$ Interpolated from \cite{Claret:2011qy}. $^{ 3 }$ Updated from \citet{Doyle:2014fk}. $^{ 4 }$ From \citet{Maxted:2011fk}.  $^{5}$ See \citet{Kipping2013}.}
\label{priors_w41}
\end{table*}
}

\section{Radial velocity data}

\onltab{
\begin{table*}
\caption{New radial velocities for WASP-47}
\begin{tabular}{lccccc}
\hline\hline
BJD -- 2\,450\,000 & RV & $\sigma_{\rm RV}$ & FWHM & Bis\\
(d) & (km\,s$^{-1}$) & (km\,s$^{-1}$) & (km\,s$^{-1}$) & (km\,s$^{-1}$)\\\hline
6205.690749 & $-27.16568$ & 0.01258 & 8.54665 & $0.01824 $\\
6213.632998 & $-27.10977$ & 0.01866 & 8.50854 & $-0.12360$\\
6246.539542 & $-27.03050$ & 0.01077 & 8.52734 & $-0.00957$\\
6269.531829 & $-27.10498$ & 0.01370 & 8.60976 & $-0.06453$\\
6270.530518 & $-26.95272$ & 0.01043 & 8.56773 & $-0.01388$\\
6437.858225 & $-27.00067$ & 0.00972 & 8.52917 & $-0.02456$\\
6452.924978 & $-27.00623$ & 0.00890 & 8.56443 & $-0.06375$\\
6459.924071 & $-27.21081$ & 0.00784 & 8.56435 & $-0.03869$\\
6477.721354 & $-27.05162$ & 0.01149 & 8.59406 & $-0.03366$\\
6486.844362 & $-26.92702$ & 0.01094 & 8.52159 & $-0.02556$\\
6505.757749 & $-27.20042$ & 0.00939 & 8.55957 & $-0.06514$\\
6512.745209 & $-26.99712$ & 0.01142 & 8.49187 & $-0.06260$\\
6514.752564 & $-27.12983$ & 0.01053 & 8.50852 & $-0.04074$\\
6515.749341 & $-26.95270$ & 0.01151 & 8.53025 & $-0.02725$\\
6530.629824 & $-27.21972$ & 0.01850 & 8.54674 & $-0.02802$\\
6534.789870 & $-27.22405$ & 0.01955 & 8.57303 & $-0.04356$\\
6559.713758 & $-27.24359$ & 0.01907 & 8.51742 & $-0.05402$\\
6590.650676 & $-26.97269$ & 0.01437 & 8.51798 & $-0.06681$\\
6616.574051 & $-27.05365$ & 0.01745 & 8.54661 & $-0.04370$\\
6782.899610 & $-26.99561$ & 0.01233 & 8.53542 & $0.00179 $\\
6808.876282 & $-27.19380$ & 0.01140 & 8.54211 & $-0.04406$\\
6831.876641 & $-26.96141$ & 0.00732 & 8.57188 & $-0.06365$\\
6859.804354 & $-27.16125$ & 0.00982 & 8.48966 & $-0.02312$\\
6885.631966 & $-26.97427$ & 0.01222 & 8.57285 & $-0.00047$\\
6915.753320 & $-26.94052$ & 0.01116 & 8.51941 & $-0.00544$\\
6950.553823 & $-27.19106$ & 0.00775 & 8.48781 & $-0.04470$\\
\multicolumn{5}{l}{\bf CORALIE upgrade}\\
7175.831586 & $-27.17962$ & 0.01397 & 8.47192 & $-0.05504$\\
7185.803516 & $-26.94267$ & 0.01691 & 8.53709 & $-0.07024$\\
7200.842220 & $-27.19041$ & 0.01242 & 8.48536 & $-0.04705$\\
7229.724940 & $-27.21735$ & 0.02664 & 8.49685 & $-0.06773$\\
7253.631858 & $-27.15060$ & 0.01096 & 8.50363 & $-0.06696$\\
7258.670969 & $-27.24158$ & 0.01503 & 8.50061 & $-0.06009$\\
\hline
\end{tabular}
\label{rv_w47}
\end{table*}
}

\onllongtab{
\begin{longtable}{lcccccc}
\caption{New radial velocities for WASP-41}\\
\hline
\hline
BJD -- 2\,450\,000 & RV & $\sigma_{\rm RV}$ & FWHM & Bis & $\log R'_{\rm HK}$ & $\sigma_{\log R'_{\rm HK}}$ \\
(d) & (km\,s$^{-1}$) & (km\,s$^{-1}$) & (km\,s$^{-1}$) & (km\,s$^{-1}$) & & \\
\hline
\endfirsthead
\caption{Continued.} \\
\hline
BJD -- 2\,450\,000 & RV & $\sigma_{\rm RV}$ & FWHM & Bis & $\log R'_{\rm HK}$ & $\sigma_{\log R'_{\rm HK}}$ \\
(d) & (km\,s$^{-1}$) & (km\,s$^{-1}$) & (km\,s$^{-1}$) & (km\,s$^{-1}$) & & \\
\hline
\endhead
\hline
\endfoot
\hline
\endlastfoot
\multicolumn{7}{l}{\bf CORALIE}\\
5577.833651 & 3.53952 & 0.01033 & 8.81258 & $-0.04113$ & & \\
5591.760168 & 3.30884 & 0.00989 & 8.80484 & $-0.01806$ & & \\
5595.802071 & 3.50882 & 0.01022 & 8.81885 & $0.00150 $& & \\
5597.776083 & 3.22930 & 0.00773 & 8.80704 & $-0.02692$ & & \\
5607.721786 & 3.41489 & 0.00873 & 8.75296 & $-0.01255$ & & \\
5621.850786 & 3.20236 & 0.00929 & 8.87451 & $-0.00206$ & & \\
5623.709905 & 3.45711 & 0.01054 & 8.81959 & $0.00608 $& & \\
5624.829879 & 3.20966 & 0.00976 & 8.80052 & $-0.00711$ & & \\
5625.826430 & 3.34101 & 0.00878 & 8.78968 & $-0.03361$ & & \\
5629.892800 & 3.45361 & 0.00973 & 8.86781 & $-0.01814$ & & \\
5635.815355 & 3.50088 & 0.01104 & 8.82368 & $-0.07613$ & & \\
5646.754274 & 3.22713 & 0.00777 & 8.79920 & $-0.01367$ & & \\
5647.630652 & 3.46375 & 0.00793 & 8.83207 & $-0.00998$ & & \\
5649.657082 & 3.19961 & 0.00800 & 8.82716 & $-0.01678$ & & \\
5675.784327 & 3.38673 & 0.01022 & 8.78597 & $0.01963 $& & \\
5676.605991 & 3.17071 & 0.00771 & 8.72380 & $-0.00364$ & & \\
5677.773655 & 3.29207 & 0.00895 & 8.72341 & $-0.04367$ & & \\
5680.715317 & 3.30325 & 0.01022 & 8.73045 & $0.00763 $& & \\
5691.683825 & 3.25986 & 0.01085 & 8.78840 & $-0.03964$ & & \\
5705.592134 & 3.43592 & 0.01549 & 8.76269 & $-0.00381$ & & \\
5716.577101 & 3.14687 & 0.00798 & 8.68982 & $-0.00490$ & & \\
5764.502263 & 3.32023 & 0.01004 & 8.60551 & $-0.00142$ & & \\
5769.525384 & 3.33383 & 0.00953 & 8.73791 & $0.03592 $& & \\
5952.825523 & 3.54587 & 0.00779 & 8.74638 & $0.00973 $& & \\
5953.844642 & 3.46188 & 0.01269 & 8.69680 & $0.00014$   & & \\
5954.786428 & 3.34226 & 0.00838 & 8.68532 & $-0.00038$ & & \\
5955.794145 & 3.57514 & 0.01921 & 8.71114 & $-0.01929$ & & \\
5958.776062 & 3.54863 & 0.00813 & 8.79174 & $-0.00790$ & & \\
5961.787850 & 3.50127 & 0.00799 & 8.77488 & $-0.00942$ & & \\
5962.724308 & 3.56518 & 0.00845 & 8.74688 & $-0.02158$ & & \\
5963.826866 & 3.33818 & 0.00795 & 8.77194 & $-0.02822$ & & \\
5974.729812 & 3.56877 & 0.00869 & 8.66496 & $-0.03499$ & & \\
5976.866742 & 3.45179 & 0.00929 & 8.76283 & $-0.01263$ & & \\
5990.888836 & 3.30386 & 0.00775 & 8.72585 & $0.00909 $& & \\
5991.827941 & 3.31816 & 0.00824 & 8.71079 & $-0.03771$ & & \\
5992.810751 & 3.56087 & 0.00741 & 8.74076 & $0.00721 $& & \\
6002.863880 & 3.35757 & 0.01043 & 8.75194 & $-0.00926$ & & \\
6003.827288 & 3.26345 & 0.01026 & 8.74074 & $0.00006 $& & \\
6004.617666 & 3.47529 & 0.01352 & 8.74425 & $-0.00198$ & & \\
6031.771885 & 3.34243 & 0.01269 & 8.65108 & $0.00612$    & & \\
6032.514592 & 3.49173 & 0.01228 & 8.67336 & $-0.01604$ & & \\
6033.583037 & 3.28047 & 0.00955 & 8.66582 & $-0.01872$ & & \\
6067.660548 & 3.16248 & 0.00874 & 8.65557 & $0.01180$   & & \\
6068.614165 & 3.36070 & 0.01015 & 8.65640 & $-0.01275$ & & \\
6113.472992 & 3.14763 & 0.01296 & 8.66391 & $-0.04187$ & & \\
6115.524431 & 3.34283 & 0.00784 & 8.62192 & $-0.02416$ & & \\
6116.552476 & 3.16081 & 0.00982 & 8.61649 & $0.01291 $& & \\
6117.563578 & 3.37099 & 0.00935 & 8.67662 & $-0.02545$ & & \\
6140.504929 & 3.17964 & 0.01699 & 8.66987 & $-0.09129$ & & \\
6150.483360 & 3.19956 & 0.01465 & 8.59768 & $-0.02625$ & & \\
6151.480248 & 3.42741 & 0.01413 & 8.62581 & $-0.00604$ & & \\
6272.851056 & 3.30816 & 0.01512 & 8.68548 & $-0.01227$ & & \\
6285.856902 & 3.46249 & 0.01426 & 8.67411 & $0.00109 $& & \\
6311.817319 & 3.26497 & 0.00849 & 8.70165 & $-0.05757$ & & \\
6314.770425 & 3.26354 & 0.00906 & 8.69916 & $-0.00225$ & & \\
6335.760287 & 3.38387 & 0.01025 & 8.69464 & $-0.03972$ & & \\
6344.893416 & 3.37439 & 0.01489 & 8.67605 & $-0.01501$ & & \\
6362.834685 & 3.50514 & 0.01487 & 8.69525 & $-0.06191$ & & \\
6364.806955 & 3.54831 & 0.01140 & 8.71556 & $-0.00035$ & & \\
6365.700329 & 3.53718 & 0.01081 & 8.74924 & $-0.02150$ & & \\
6372.694569 & 3.30787 & 0.01324 & 8.71766 & $-0.00158$ & & \\
6401.508276 & 3.52006 & 0.00950 & 8.71947 & $-0.02537$ & & \\
6419.727309 & 3.46173 & 0.00981 & 8.62232 & $0.01914 $& & \\
6427.640513 & 3.25900 & 0.01451 & 8.64472 & $-0.02081$ & & \\
6441.643745 & 3.48042 & 0.01110 & 8.64098 & $-0.01737$ & & \\
6450.560548 & 3.46452 & 0.00870 & 8.65585 & $-0.02942$ & & \\
6484.570287 & 3.42376 & 0.01376 & 8.57077 & $-0.01220$ & & \\
6496.467235 & 3.44219 & 0.00958 & 8.62008 & $-0.01283$ & & \\
6506.482624 & 3.27505 & 0.01018 & 8.59371 & $-0.00833$ & & \\
6514.470015 & 3.38330 & 0.00963 & 8.64153 & $-0.02157$ & & \\
6646.830447 & 3.29643 & 0.00838 & 8.70805 & $-0.02489$ & & \\
6685.792726 & 3.46022 & 0.00978 & 8.73929 & $-0.03507$ & & \\
6699.747165 & 3.22170 & 0.00926 & 8.76697 & $-0.03725$ & & \\
6713.750758 & 3.42864 & 0.00862 & 8.70401 & $-0.03290$ & & \\
6723.823895 & 3.25117 & 0.00909 & 8.65259 & $-0.01733$ & & \\
6738.590042 & 3.36799 & 0.00905 & 8.64089 & $-0.04485$ & & \\
6748.823614 & 3.36232 & 0.00913 & 8.71895 & $-0.04756$ & & \\
6830.613737 & 3.26410 & 0.01257 & 8.63168 & $-0.05510$ & & \\
\multicolumn{7}{l}{\bf HARPS}\\
5655.551700 & 3.22759 & 0.00363 & 7.59140 & $-0.00330$ & $-4.4430$ & 0.0172\\
5655.853857 & 3.23851 & 0.00633 & 7.58522 & $0.00779$    & $-4.4359$ & 0.0332\\
5656.536367 & 3.39812 & 0.00321 & 7.61857 & $0.02087$    & $-4.4596$ & 0.0154\\
5656.697723 & 3.43224 & 0.00265 & 7.57389 & $0.00204$    & $-4.4619$ & 0.0116\\
5656.862517 & 3.45792 & 0.00341 & 7.59920 & $0.02669$    & $-4.4865$ & 0.0194\\
5681.502346 & 3.47325 & 0.00470 & 7.54826 & $-0.01900$ & $-4.4496$ & 0.0228\\
5681.747792 & 3.46028 & 0.00331 & 7.52105 & $-0.02056$ & $-4.5356$ & 0.0190\\
5682.508177 & 3.27667 & 0.00278 & 7.54712 & $-0.01070$ & $-4.4821$ & 0.0126\\
5682.746435 & 3.22961 & 0.00357 & 7.55328 & $-0.00814$ & $-4.4868$ & 0.0178\\
5683.607603 & 3.29092 & 0.00271 & 7.53416 & $-0.00548$ & $-4.4662$ & 0.0116\\
5684.631995 & 3.48142 & 0.00302 & 7.56635 & $-0.01153$ & $-4.4767$ & 0.0136\\
5685.587048 & 3.27317 & 0.00249 & 7.57172 & $0.00787$    & $-4.4737$ & 0.0104\\
5711.539738 & 3.39743 & 0.00428 & 7.55737 & $0.00110$    & $-4.4846$ & 0.0208\\
5712.569788 & 3.38172 & 0.00379 & 7.55858 & $-0.00532$ & $-4.5048$ & 0.0191\\
5716.557913 & 3.16014 & 0.00336 & 7.47622 & $0.01511$    & $-4.6466$ & 0.0252\\
5728.497874 & 3.20106 & 0.00352 & 7.48383 & $-0.00680$ & $-4.5084$ & 0.0169\\
5729.498999 & 3.27894 & 0.00418 & 7.50356 & $-0.00211$ & $-4.4838$ & 0.0202\\
5749.508251 & 3.28400 & 0.00444 & 7.47420 & $-0.01793$ & $-4.5368$ & 0.0250\\
5753.559500 & 3.19748 & 0.01026 & 7.50218 & $-0.05764$ & $-4.4422$ & 0.0560\\
5754.523986 & 3.43646 & 0.00761 & 7.50424 & $-0.02168$ & $-4.4780$ & 0.0417\\
5776.464239 & 3.41784 & 0.00841 & 7.49327 & $0.00614$    & $-4.4815$ & 0.0489\\
6020.679485 & 3.49349 & 0.00572 & 7.45429 & $-0.01218$ & $-4.5778$ & 0.0433\\
6021.672381 & 3.25952 & 0.01203 & 7.46629 & $-0.05230$ & $-4.4977$ & 0.0799\\
6029.682164 & 3.47547 & 0.00966 & 7.45214 & $-0.03355$ & $-4.5089$ & 0.0648\\
6058.586834 & 3.17781 & 0.01144 & 7.38666 & $-0.03210$ & $-4.4681$ & 0.0662\\
6059.587695 & 3.43300 & 0.01062 & 7.46955 & $0.01536$    & $-4.5694$ & 0.1073\\
6092.564596 & 3.24052 & 0.00991 & 7.47168 & $-0.02810$ & $-5.0597$ & 0.2935\\
6093.534543 & 3.45736 & 0.00745 & 7.45766 & $-0.00441$ & $-4.5485$ & 0.0567\\
6094.550435 & 3.26252 & 0.01267 & 7.42991 & $-0.03793$ & $-4.5425$ & 0.1083\\
6151.471858 & 3.43600 & 0.01277 & 7.41152 & $0.00263$    & $-4.5729$ & 0.1182\\
\multicolumn{7}{l}{\bf HARPS RM sequence}\\
5654.541742 & 3.44100 & 0.00408 & 7.58857 & $-0.01057$ & $-4.4695$ & 0.0211\\
5654.627495 & 3.42446 & 0.00309 & 7.56772 & $-0.00340$ & $-4.4528$ & 0.0134\\
5654.694950 & 3.40718 & 0.00521 & 7.56499 & $-0.03342$ & $-4.4779$ & 0.0331\\
5654.701224 & 3.40040 & 0.00512 & 7.55081 & $0.00192$  & $-4.5322$ & 0.0367\\
5654.707323 & 3.39846 & 0.00446 & 7.57235 & $-0.02608$ & $-4.4858$ & 0.0272\\
5654.713365 & 3.39405 & 0.00544 & 7.54723 & $0.00574$  & $-4.5201$ & 0.0379\\
5654.719580 & 3.39300 & 0.00525 & 7.58325 & $-0.02580$ & $-4.5016$ & 0.0384\\
5654.725854 & 3.39530 & 0.00487 & 7.57183 & $-0.00835$ & $-4.4374$ & 0.0310\\
5654.731722 & 3.39696 & 0.00480 & 7.57224 & $0.01508$  & $-4.4085$ & 0.0283\\
5654.738169 & 3.38889 & 0.00486 & 7.58768 & $-0.00323$ & $-4.5086$ & 0.0356\\
5654.744152 & 3.38091 & 0.00449 & 7.57718 & $-0.03317$ & $-4.5097$ & 0.0331\\
5654.750368 & 3.38292 & 0.00460 & 7.62232 & $0.01020$  & $-4.5330$ & 0.0359\\
5654.756525 & 3.38866 & 0.00483 & 7.58194 & $-0.01348$ & $-4.4997$ & 0.0358\\
5654.762741 & 3.39363 & 0.00506 & 7.57987 & $-0.01364$ & $-4.4967$ & 0.0373\\
5654.768667 & 3.40373 & 0.00531 & 7.55134 & $0.00864$  & $-4.4188$ & 0.0320\\
5654.774824 & 3.41298 & 0.00579 & 7.56770 & $-0.01484$ & $-4.4786$ & 0.0416\\
5654.781317 & 3.40804 & 0.00595 & 7.55392 & $-0.01020$ & $-4.4952$ & 0.0432\\
5654.787243 & 3.38972 & 0.00548 & 7.58143 & $-0.01048$ & $-4.5308$ & 0.0446\\
5654.793586 & 3.38582 & 0.00517 & 7.57592 & $-0.03576$ & $-4.4906$ & 0.0375\\
5654.799339 & 3.38918 & 0.00622 & 7.58407 & $-0.01323$ & $-4.4909$ & 0.0470\\
5654.805785 & 3.36643 & 0.00549 & 7.60594 & $-0.00493$ & $-4.5604$ & 0.0485\\
5654.811885 & 3.36490 & 0.00604 & 7.61466 & $0.00560$  & $-4.4852$ & 0.0462\\
5654.818043 & 3.35075 & 0.00696 & 7.61693 & $-0.02194$ & $-4.4819$ & 0.0523\\
5654.824385 & 3.34658 & 0.00659 & 7.62860 & $-0.00536$ & $-4.5539$ & 0.0568\\
5654.835242 & 3.33034 & 0.00638 & 7.60329 & $-0.00396$ & $-4.5984$ & 0.0655\\
5654.841399 & 3.32018 & 0.00596 & 7.56792 & $0.01640$  & $-4.6577$ & 0.0683\\
5654.847557 & 3.31924 & 0.00596 & 7.60258 & $-0.03724$ & $-4.5498$ & 0.0518\\
5654.853541 & 3.33239 & 0.00702 & 7.58480 & $0.02337$  & $-4.6217$ & 0.0726\\
5654.859930 & 3.34743 & 0.00778 & 7.56688 & $-0.02332$ & $-4.4810$ & 0.0624\\
5654.866029 & 3.34820 & 0.00757 & 7.61977 & $0.01059$  & $-4.6210$ & 0.0795\\
5654.872071 & 3.34226 & 0.00772 & 7.60082 & $-0.03158$ & $-4.5566$ & 0.0707\\
5654.878402 & 3.33363 & 0.00903 & 7.59843 & $0.00049$  & $-4.7133$ & 0.1221\\
5654.884502 & 3.33904 & 0.00873 & 7.57028 & $0.00809$  & $-4.5821$ & 0.0908\\
5654.890659 & 3.34222 & 0.00860 & 7.59899 & $-0.03549$ & $-4.3253$ & 0.0494\\
5654.896829 & 3.34804 & 0.00859 & 7.57287 & $0.01621$  & $-4.5036$ & 0.0757\\
\end{longtable}
\label{rv_w41}
}

\end{appendix}

\end{document}